\documentclass[usenatbib]{mnras}

\usepackage{newtxtext,newtxmath}

\usepackage[T1]{fontenc}

\DeclareRobustCommand{\VAN}[3]{#2}
\let\VANthebibliography\thebibliography
\def\thebibliography{\DeclareRobustCommand{\VAN}[3]{##3}\VANthebibliography}

\usepackage{graphicx}	
\usepackage{amsmath}	
\usepackage{booktabs}

\title[Matched filters for precision cosmology]{Understanding matched filters for precision cosmology}

\author[\'{I}. Zubeldia et al.]{
\'{I}\~{n}igo Zubeldia,$^{1}$\thanks{E-mail: inigo.zubeldia@manchester.ac.uk}
Aditya Rotti,$^{1}$
Jens Chluba$^{1}$
and Richard Battye$^{1}$
\\
$^{1}$Jodrell Bank Centre for Astrophysics, University of Manchester, Manchester M13 9PL UK
}

\date{Accepted XXX. Received YYY; in original form ZZZ}

\pubyear{2015}

\begin{document}
\label{firstpage}
\pagerange{\pageref{firstpage}--\pageref{lastpage}}
\maketitle

\begin{abstract}
Matched filters are routinely used in cosmology in order to detect galaxy clusters from mm observations through their thermal Sunyaev-Zeldovich (tSZ) signature. In addition, they naturally provide an observable, the detection signal-to-noise or significance, which can be used as a mass proxy in number counts analyses of tSZ-selected cluster samples. In this work, we show that this observable is, in general, non-Gaussian, and that it suffers from a positive bias, which we refer to as optimisation bias. Both aspects arise from the fact that the signal-to-noise is constructed through an optimisation operation on noisy data, and hold even if the cluster signal is modelled perfectly well, no foregrounds are present, and the noise is Gaussian. After reviewing the general mathematical formalism underlying matched filters, we study the statistics of the signal-to-noise with a set Monte Carlo mock observations, finding it to be well-described by a unit-variance Gaussian for signal-to-noise values of $6$ and above, and quantify the magnitude of the optimisation bias, for which we give an approximate expression that may be used in practice. We also consider the impact of the bias on the cluster number counts of \textit{Planck} and the Simons Observatory (SO), finding it to be negligible for the former and potentially significant for the latter.
\end{abstract}

\begin{keywords}
galaxies: clusters: general -- cosmology: observations -- cosmology: diffuse radiation
\end{keywords}



\section{Introduction}\label{sec:intro}

First proposed in the galaxy cluster context more than two decades ago \citep{Haehnelt1996,Herranz2002,Melin2006}, matched filters have become a standard tool with which to detect and characterise galaxy clusters from CMB observations via their thermal Sunyaev-Zeldovich (tSZ) signature. Indeed, matched filters have been used in the construction of the cluster catalogues derived from SPT, ACT, and \textit{Planck} data (e.g., \citealt{Vanderlinde2010,Hasselfield2013,Planck2014,Bleem2015,Planck2016xxvii,Hilton2020,Melin2021}), and as a result they have also played an important role in the cosmological analyses based on these cluster samples (e.g., \citealt{Vanderlinde2010,Hasselfield2013,Planck2014XX,Bleem2015,Ade2016,Bocquet2018,Zubeldia2019}).

Matched filters provide a natural quantity, the detection significance or signal-to-noise, that can be used to construct cluster samples by selecting the clusters whose signal-to-noise falls above a certain threshold (e.g., \citealt{Planck2016xxvii,Hilton2020}). The signal-to-noise can also be employed as a mass proxy in a cluster number counts analysis, in which the cluster abundance across a number of observables (usually, redshift and one or more `mass proxies', that is, observables that scale with cluster mass) is used to constrain cosmology (e.g., \citealt{Ade2016,Bocquet2018,Zubeldia2019}; see \citealt{Allen2011} for a review). Essential for the success of such analyses is the accurate determination of the mass--observable(s) scaling relation(s), which allows to link the predicted cluster abundance across mass and redshift, as given by the halo mass function, to the observed abundance. At present, this constitutes the main source of systematic uncertainty in cluster number counts analyses, and, as such, a lot of effort is being devoted to the calibration of mass--observable relations (e.g., \citealt{Nicola2020,Zubeldia2020,Andrade-Santos2021}; see \citealt{Pratt2019} for a review). Any significant biases in any of the mass--observable relation could lead to biased cosmological constraints. Thus, if the matched filter signal-to-noise is to be used as a mass observable in a number counts analysis, its relation to cluster mass and redshift must be well understood.

Relating the cluster mass and redshift to the observed signal-to-noise is usually thought of as a two-step process (see, e.g., \citealt{Ade2016} and \citealt{Bocquet2018}). First, the cluster mass and redshift are related to some mean signal-to-noise via a set of scaling relations, which after the addition of some `intrinsic scatter' (due to cluster triaxiality, deviations from the mean profile, etc.) one obtains the `true' signal-to-noise of a cluster, i.e., the signal-to-noise that would be measured after ensemble averaging over what is understood as noise in the matched filtering process. Then, a layer of additive `noise' or `observational scatter' connects the true signal-to-noise with the observed one. This observational scatter is typically assumed to be Gaussian-distributed and to have a standard deviation equal to unity and a mean equal to either the true-signal-to-noise (\textit{Planck} and ACT; see, e.g., \citealt{Hasselfield2013,Ade2016}) or a corrected version of it (SPT; see, e.g., \citealt{Bocquet2018}).

In this work we focus on this second layer of the scatter, which is comparatively much less studied than the intrinsic scatter. In order to do this, we first review the matched filter formalism, deriving the matched filter using a maximum-likelihood approach. We show that, as opposed to what is typically assumed, the observational scatter is, in general, non-Gaussian. We also show that the observed signal-to-noise is, in general, biased high with respect to the true signal-to-noise. Both the non-Gaussianity and the bias are a result of the signal-to-noise being obtained through an optimisation process on noisy data. As a consequence, they hold even if the signal is modelled perfectly, no foregrounds are present, and the noise in the data is Gaussian, which is indeed the case we consider. After presenting these theoretical considerations, we then use a set of Monte Carlo (MC) mock observations in order to quantify the non-Gaussianity of the observational scatter, its variance, and the magnitude of the optimisation bias. We find that for signal-to-noise values of $\gtrsim 6$, the scatter can taken to be approximately Gaussian and with a variance equal to unity. We also find that, for a similar signal-to-noise range, the optimisation bias can be described accurately with a simple approximate expression. This expression is the same as the correction used in the SPT analyses \citep{Bocquet2018}, which was originally introduced in \citet{Vanderlinde2010} and justified with a rather heuristic argument. Here, we provide further insight into its origin and a more detailed study of its validity in the Gaussian noise regime. Finally, we quantify the impact of the optimisation bias in the cluster number counts of \textit{Planck} and the Simons Observatory (SO; \citealt{SO2019}), finding it to be negligible for the former and potentially significant for the latter.

This paper is organised as follows. In Section \ref{sec:theory} we review the matched filter formalism, deriving the matched filter from a maximum likelihood point of view (Section \ref{sec:mf}) and noting the biases that arise from this process, both in the retrieved parameter estimates (Section \ref{sec:solutionbias}) and in the signal-to-noise (the `optimisation bias'; Sections \ref{sec:optbias} and \ref{sec:partialopt}). Then, after a brief review of how matched filters are used in practice in the tSZ context (Section \ref{sec:context}), in Section \ref{mc_test} we use a set of MC mock observations in order to study the statistics of the observational scatter of the signal-to-noise, quantifying its departure from Gaussianity, its variance, and its mean (the latter as quantified through the optimisation bias). We also comment briefly on other possible biases in the signal-to-noise, such as that arising from estimating the noise covariance from the data in a naive way (Section \ref{sec:cov}). Next, in Section \ref{planck} we consider the impact of the optimisation bias on the cluster number counts of \textit{Planck} and SO, and finally conclude in Section \ref{sec:conclusion}.




\vspace{-3mm}
\section{Matched filtering: a likelihood approach}\label{sec:theory}

\subsection{Matched filtering as maximum likelihood estimation}\label{sec:mf}

Consider a data vector $\mathbfit{d}$. This can be, e.g., a map of the Compton-$y$ parameter, a map of the lensing convergence, or a set of intensity maps across several frequencies. In the first two cases, the data vector dimension is equal to the number of pixels, $n$, whereas in the third it is equal to the number of pixels times the number of frequency channels. Let us assume that this data vector can be written as

\begin{equation}\label{model}
    \mathbfit{d} = s_0 \mathbfit{s} (\bmath{\theta}) + \mathbfit{n}.
\end{equation}
Here, $s_0 \mathbfit{s} (\bmath{\theta})$ is the signal that is present in the data (e.g., the Compton-$y$ signal due to a galaxy cluster), which we assume that can be written as an amplitude, or normalisation, parameter $s_0$ times a template function $\mathbfit{s} (\bmath{\theta})$, which depends upon a number of $f$ additional parameters $\bmath{\theta}$. This set of parameters can include, for example, a size parameter (e.g., $\theta_{500}$ for a galaxy cluster), as well as location parameters specifying the position of the peak of the signal in the map. In addition, our data vector has some additive noise, $\mathbfit{n}$, which we will assume to be Gaussian distributed and with zero mean. In practice, this noise term needs not be instrumental noise alone: it can also include the signal from one, or several, foregrounds, as long as it is Gaussian distributed and has zero mean. The logarithm of the likelihood (hereafter, the log-likelihood) of our data can then be written as

\begin{align}
\label{likelihood}
    \ln \mathcal{L} (s_0,\bmath{\theta})
    &= - \frac{1}{2} [\mathbfit{d} - s_0 \mathbfit{s} (\bmath{\theta})]^{T} \mathbfss{C}^{-1} [\mathbfit{d} - s_0 \mathbfit{s} (\bmath{\theta})]
    \nonumber\\
    &\qquad\qquad\qquad
    - \frac{1}{2} \ln \left[ (2 \upi)^n \mathrm{det}(\mathbfss{C})\right],
\end{align}
where $\mathbfss{C}$ is the noise covariance, which we assume to be known. The second term of this likelihood is independent from the parameters, and so we will ignore it in the following.

Let us now consider the set of parameter values that maximises the log-likelihood, $\left\{\hat{s}_0,\hat{\bmath{\theta}} \right\}$. We first optimise the log-likelihood with respect to $s_0$. Imposing $\partial \ln \mathcal{L} / \partial s_0 = 0$ yields

\begin{equation}\label{s0}
    \hat{s}_0 (\bmath{\theta}) = \frac{ \mathbfit{s} (\bmath{\theta})^T \mathbfss{C}^{-1} \mathbfit{d} }{\mathbfit{s} (\bmath{\theta})^T \mathbfss{C}^{-1} \mathbfit{s} (\bmath{\theta})},
\end{equation}
which is the well-known matched filter estimate of $s_0$ at a given value of $\bmath{\theta}$, and which has an associated standard deviation $\sigma_{s_0}$ given by

\begin{equation}\label{noise}
    \sigma_{s_0} (\bmath{\theta}) = \left[\mathbfit{s} (\bmath{\theta})^T \mathbfss{C}^{-1} \mathbfit{s} (\bmath{\theta}) \right]^{-1/2}.
\end{equation}
We can then define the signal-to-noise ratio of the detection as

\begin{equation}\label{snr}
    q (\bmath{\theta}) \equiv \frac{\hat{s}_0 (\bmath{\theta})}{\sigma_{s_0} (\bmath{\theta})} = \frac{ \mathbfit{s} (\bmath{\theta})^T \mathbfss{C}^{-1} \mathbfit{d}  }{\left[ \mathbfit{s} (\bmath{\theta})^T \mathbfss{C}^{-1} \mathbfit{s} (\bmath{\theta}) \right]^{1/2}},
\end{equation}
where we have stressed its dependence on $\bmath{\theta}$ by leaving it explicit.

By substituting the expression for $\hat{s}_0$ of Eq. (\ref{s0}) into Eq. (\ref{likelihood}), the log-likelihood can be rewritten as

\begin{equation}\label{likelihood2}
    \ln \mathcal{L} (\bmath{\theta}) = -\frac{1}{2} \mathbfit{d}^T \mathbfss{C}^{-1} \mathbfit{d} + \frac{1}{2} q (\bmath{\theta})^2.
\end{equation}
We note that the likelihood no longer depends on $s_0$, as this parameter has already been maximised over, and we recall that we have dropped the second, parameter-independent term. From Eq. (\ref{likelihood2}) it follows that the set of parameter values $\bmath{\theta}$ that maximises the log-likelihood is equal to that which maximises the signal-to-noise $q$. That is, the maximum likelihood solution is equal to the value of $\bmath{\theta}$ that maximises the matched filter signal-to-noise, along with the corresponding matched filter amplitude estimate, $\hat{s}_0({\hat{\bmath{\theta}}})$.

When matched filters are used in practice, it is customary to maximise the signal-to-noise with respect to $\bmath{\theta}$ (e.g., \citealt{Melin2006}). Here, we have shown that this \textit{matched filter solution} is equal to the maximum likelihood solution. In brief, matched filtering with signal-to-noise maximisation can be understood as maximum likelihood estimation for a Gaussian likelihood and a model which can be written as an amplitude parameter times a template function.

\subsection{Matched filter solution bias}\label{sec:solutionbias}

By identifying the matched filter solution with the maximum likelihood solution, $\left\{ \hat{s}_0,\hat{\bmath{\theta}} \right\}$, it follows that the matched filter estimates of $s_0$ and $\bmath{\theta}$ are, in general, biased, as they are in general maximum likelihood estimators. We thus warn against using matched filter solutions without proper consideration of this bias. As we illustrate numerically in Section \ref{sec:paramest}, the magnitude of the bias typically decreases with the signal-to-noise, being arbitrarily small for arbitrarily large values of signal-to-noise (i.e., for arbitrarily low noise). Care should be taken, however, when conducting inference from the combination of the matched filter solutions of a set of different observations, i.e., a set of $m$ data vectors $\mathbfit{d}$. In this case, the overall signal-to-noise, which will set the precision of the inference, will be larger than the signal-to-noise of the individual measurements, increasing as $\sqrt{m}$, but the bias will not decrease accordingly.

We also note that the bias is parametrisation-dependent. As an illustration, consider, for example, that $\bmath{\theta}$ includes two location parameters, $\theta_x$ and $\theta_y$, which specify the Cartesian coordinates of the peak of the signal. By symmetry, their matched filter solution, $\hat{\theta}_x$ and $\hat{\theta}_y$, cannot be biased, i.e., $\left\langle \hat{\theta}_x \right\rangle = \theta_x^{\mathrm{true}}$ and $\left\langle \hat{\theta}_y \right\rangle = \theta_y^{\mathrm{true}}$, where angular brackets denote ensemble averaging over noise realisations. Consider now an alternative and fully equivalent parametrisation, $\theta = (\theta_x^2+\theta_y^2)^{1/2}$ and $\phi = \tan^{-1} (\theta_y/\theta_x)$. It should be apparent that the matched filter solution for $\theta$ is biased, as its expected value receives contributions from the second-order moments of $\hat{\theta}_x$ and $\hat{\theta}_y$. We illustrate this point numerically in Section \ref{sec:paramest}.

\subsection{Using the optimal $q$: optimisation bias}\label{sec:optbias}

In some applications of matched filtering (e.g., galaxy cluster detection in tSZ surveys, see Section \ref{sec:cluster}), the main quantity of interest is the signal-to-noise ratio, $q$, and, more specifically, its maximum value, that is, the value that it takes at the maximum likelihood solution, $q (\hat{\bmath{\theta}})$. We refer to this quantity as the `optimal signal-to-noise' and denote it with $q_{\mathrm{opt}}$. In order to be able to infer information from a measured value of $q_{\mathrm{opt}}$, one needs to be able to relate this quantity to the \textit{true mean} signal-to-noise ratio, which is given by ensemble-averaging Eq. (\ref{snr}) over noise realisations at the true parameter values, and which we denote with $\bar{q}_{\mathrm{t}}$. That is, one needs to determine $P(q_{\mathrm{opt}}|\bar{q}_{\mathrm{t}})$. This is usually known as `observational scatter' or `noise' (e.g., \citealt{Zubeldia2019}), as opposed to the intrinsic scatter due to triaxiality, inhomogeneities, etc., which we ignore in this work.

For a given noise realisation $\mathbfit{n}$, $P(q_{\mathrm{opt}} | \bar{q}_{\mathrm{t}}, \mathbfit{n})$ is a $\delta$ function centred at $q (\hat{\bmath{\theta}} (\bmath{\theta}_{\mathrm{t}}, \mathbfit{n}), \mathbfit{n})$, where we have written explicitly the dependence of the matched filter solution $\hat{\bmath{\theta}}$ on the true parameter values and on the noise. Hence, we can write

\begin{equation}
    P(q_{\mathrm{opt}}|\bar{q}_{\mathrm{t}}) = \int \delta \left[q_{\mathrm{opt}} - q(\hat{\bmath{\theta}} (\bmath{\theta}_{\mathrm{t}}, \mathbfit{n}), \mathbfit{n}) \right]  P (\mathbfit{n}) d^n \mathbfit{n},
\end{equation}
where $P (\mathbfit{n})$ is the noise probability density function, which, if the noise is Gaussian, as we assume, is a multi-variate Gaussian. This is, in general, a non-Gaussian distribution, reflecting the fact that optimisation is a non-linear operation. Intuitively, to the Gaussian variation due to the noise one should add the variation due to evaluating Eq. (\ref{snr}) at different values of $\bmath{\theta}$ (for each measurement, its matched filter solution $\hat{\bmath{\theta}}$). The $m$-th moment of this distribution can be written as

\begin{equation}\label{moments}
   \langle q_{\mathrm{opt}}^m \rangle = \int q(\hat{\bmath{\theta}} (\bmath{\theta}_{\mathrm{t}}, \mathbfit{n}))^m   P (\mathbfit{n})  d^n \mathbfit{n}.
\end{equation}
This is a set of high-dimensional integrals which are difficult to compute in practice. However, their values can be estimated in a fast way with Monte Carlo (MC) simulations, as we do in Section \ref{sec:cluster} for $m=1, \dots, 4$ in the context of galaxy cluster tSZ observations. Nevertheless, further insight into the value of the first moment, the mean, which we denote with $\bar{q}_{\mathrm{opt}}$, can be gained with the following argument.



Consider the value of $-2 \ln \mathcal{L}$, i.e., of the sum of the residuals. If we evaluate it at the true value of $\bmath{\theta}$, $\bmath{\theta}_{\mathrm{t}}$, and at the corresponding matched filter value of $s_0$, $\hat{s}_0 (\bmath{\theta}_{\mathrm{t}})$, there is only one fitting parameter, $s_0$, the model being linear in it. Thus, in this case $-2 \ln \mathcal{L}$ is $\chi^2$-distributed with $n - 1$ degrees of freedom, where we recall that $n$ is the dimension of our data vector (e.g., \citealt{Andrae2010}). Using Eq. (\ref{likelihood2}), its expected value can be written as

\begin{equation}
\left\langle -2 \ln \mathcal{L} (\hat{s}_0 (\bmath{\theta}_{\mathrm{t}}),\bmath{\theta}_{\mathrm{t}}) \right\rangle = \mathbfit{t}^T \mathbfss{C}^{-1} \mathbfit{t} + n  -  \langle q (\bmath{\theta}_{\mathrm{t}})^2\rangle = n-1,
\end{equation}
where $\mathbfit{t}$ it the true signal (i.e., the model evaluated at the true parameter values), and where we have ignored the second, parameter-independent term. Since $\langle q (\bmath{\theta}_{\mathrm{t}}) \rangle = \bar{q}_{\mathrm{t}}$, and since $q (\bmath{\theta}_{\mathrm{t}})$ has unit variance, it follows that

\begin{equation}\label{chi1}
\mathbfit{t}^T \mathbfss{C}^{-1} \mathbfit{t} + n - 1 - \bar{q}_{\mathrm{t}}^2 = n-1.
\end{equation}
On the other hand, $-2 \ln \mathcal{L}$ is not $\chi^2$-distributed if we evaluate it for the matched filter solution, in which $f$ additional parameters are fitted, as now the model is, in general, non-linear in the parameters (see, e.g., \citealt{Andrae2010}). Equivalently, this can be seen as a consequence of $q_{\mathrm{opt}}$ being non-Gaussian. This means, in particular, that we cannot compute its expected value by invoking the number of degrees of freedom in the fit, as we did in the previous case. However, this is still a well-defined quantity, which we write as $n-1-f_{\mathrm{eff}}$, where we have introduced $f_{\mathrm{eff}}$ as the `effective number of additional fitting parameters'. The value of $f_{\mathrm{eff}}$ cannot be determined a priori, but it is clear that $f_{\mathrm{eff}} \ge 0$. We can now write

\begin{equation}
\left\langle -2 \ln \mathcal{L} (\hat{s}_0 ( \hat{\bmath{\theta}}),\bmath{\hat{ \theta}}) \right\rangle = \mathbfit{t}^T \mathbfss{C}^{-1} \mathbfit{t} + n  -  \langle q_{\mathrm{opt}}^2\rangle = n-1-f_{\mathrm{eff}},
\end{equation}
and so

\begin{equation}\label{chi2}
\mathbfit{t}^T \mathbfss{C}^{-1} \mathbfit{t} + n - \sigma_{\mathrm{opt}}^2 - \bar{q}_{\mathrm{opt}}^2 = n-1-f_{\mathrm{eff}},
\end{equation}
where $\sigma_{\mathrm{opt}}$ denotes the standard deviation of $q_{\mathrm{opt}}$, which need not be equal to unity, as in the Gaussian case. Subtracting Eq. (\ref{chi1}) from Eq. (\ref{chi2}) and rearranging, we find

\begin{equation}\label{correction}
\bar{q}_{\mathrm{opt}} = \sqrt{ \bar{q}_{\mathrm{t}}^2 + f_{\mathrm{eff}} + \Delta {\sigma^2} },
\end{equation}
where $\Delta {\sigma^2} \equiv 1 - \sigma_{\mathrm{opt}}^2 $.

Thus, $\bar{q}_{\mathrm{opt}}$ is biased with respect to $\bar{q}_{\mathrm{t}}$. We refer to this bias as \textit{optimisation bias}, and note that it is a positive bias, with $\bar{q}_{\mathrm{opt}} > \bar{q}_{\mathrm{t}}$. This can in fact be deduced from a simple argument: for any noise realisation, $q_{\mathrm{opt}} > q (\bmath{\theta}_{\mathrm{t}})$ by construction (except in the case in which $\hat{\bmath{\theta}} =  \bmath{\theta}_{\mathrm{t}}$, which happens with formally zero probability), and thus $\bar{q}_{\mathrm{opt}} > \bar{q}_{\mathrm{t}}$. Intuitively: by allowing for additional freedom in the fit, the signal-to-noise is boosted, as one is effectively fitting noise.


In summary, $P(q_{\mathrm{opt}}|\bar{q}_{\mathrm{t}})$ is a non-Gaussian distribution with moments given by Eq. (\ref{moments}) and whose mean can be also written as in Eq. (\ref{correction}). In order for a set of $q_{\mathrm{opt}}$ measurements to be used in practice, one has to determine how much $P(q_{\mathrm{opt}}|\bar{q}_{\mathrm{t}})$ deviates from Gaussianity, as well as the values of $f_{\mathrm{eff}}$ and $\sigma_{\mathrm{opt}}$, both of which can be a function of $\bar{q}_{\mathrm{t}}$. In Section \ref{sec:cluster} we do so using MC simulations for the case in which the signal to be extracted is the tSZ signal from a galaxy cluster, finding that, for sufficiently large values of $\bar{q}_{\mathrm{t}}$ ($\simeq 6$), $P(q_{\mathrm{opt}}|\bar{q}_{\mathrm{t}})$ approaches a unit-variance Gaussian and $f_{\mathrm{eff}}$ approaches the actual number of additional fitting parameters, $f$.

\subsection{Partial optimisation}\label{sec:partialopt}

Thus far we have assumed that the optimisation procedure finds the true signal-to-noise maximum, i.e., that the true maximum likelihood solution is obtained. In some applications of matched filtering, however, the signal-to-noise is evaluated on a given parameter grid, with the `matched filter solution' and its associated signal-to-noise, $q_{\mathrm{opt}}$, now being the grid coordinates at which the signal-to-noise is maximised and the corresponding signal-to-noise value, respectively. As we will see in Section \ref{sec:cluster}, this \textit{partial} optimisation can have an impact on the statistics of $q_\mathrm{\mathrm{opt}}$. In general, a dependency on the grid chosen is introduced, with the statistics of the partially-optimised $q_\mathrm{\mathrm{opt}}$ being obviously expected to approach those of the \textit{true} $q_\mathrm{\mathrm{opt}}$ as the grid spacing is decreased. As we illustrate in Section \ref{sec:cluster}, the impact of this partial optimisation must be carefully considered if unbiased information is to be extracted from partially-optimised $q_\mathrm{\mathrm{opt}}$ measurements.



\section{A practical application: SZ galaxy cluster detection and number counts}\label{sec:cluster}

\subsection{Matched filtering in the galaxy cluster context}\label{sec:context}

As noted in the Introduction, matched filters are now routinely used in order to detect and characterise galaxy clusters from mm data. They have indeed been used in experiments such as SPT, ACT, and \textit{Planck} (\citealt{Vanderlinde2010,Hasselfield2013,Planck2014,Bleem2015,Planck2016xxvii,Hilton2020,Melin2021}), and will also continue to be used in upcoming experiments such as SO and CMB-S4 \citep{Abazajian2016,SO2019}. In these works, galaxy clusters are detected following, roughly, the same general procedure. First, the cluster search sky area is divided into a set of patches, which allows for local noise estimation. Then, a model for the cluster tSZ signal is assumed. This model is parametrised as our signal in Eq. (\ref{model}), with an overall amplitude parameter and a template function, which is typically taken to depend on two sky location parameters and on an angular size parameter. For each of the patches, this template is used to construct a matched filter, for which the signal-to-noise (in our notation, $q$) is evaluated across a number of sky locations (usually, for each pixel) and angular sizes. We note that if observations at more than one frequency channel are used, the matched filter operates across the maps at the different frequencies considered, and is usually referred to as a multifrequency matched filter (MMF). If we put the frequency and spatial indices on the same footing, however, an MMF can be seen as a `single-frequency' matched filter, which means that the formalism and the insights presented in Section \ref{sec:theory} remain entirely valid.

For each patch, the sky locations of the cluster candidates in it are then obtained from the local peaks in the signal-to-noise map, with the specific procedure varying from analysis to analysis, but being in essence a maximisation of the signal-to-noise over sky location (a `partial' maximisation, however; see Section \ref{sec:partialopt}). The characterisation of the cluster signal, nonetheless, depends more significantly on the analysis considered. In \textit{Planck} and SPT, the signal-to-noise (partially) maximised over sky location and angular size is used ($q_{\mathrm{opt}}$ in our notation; e.g., \citealt{Vanderlinde2010,Planck2014}). In ACT, on the other hand, the signal-to-noise is maximised only to assign a sky location, the matched filter being only used at a single fixed angular size, at which the signal amplitude ($\hat{s}_0$ in our notation) is computed and used as the cluster observable. In the likelihood, however, this $\hat{s}_0$ measurement has as associated standard deviation the corresponding matched filter standard deviation (given by Eq. \ref{noise}), so using $\hat{s}_0$ at a fixed angular scale as an observable is equivalent to using the corresponding $q$ measurement. There is an obvious loss of statistical power due to fixing the angular scale, but, as argued in \citet{Hasselfield2013} and \citet{Hilton2020}, doing this should partially mitigate against what we understand is the optimisation bias.

In short, as the cluster tSZ observable, SPT and \textit{Planck} use $q_{\mathrm{opt}}$ with (partial) optimisation over sky location and angular size, whereas ACT effectively uses $q_{\mathrm{opt}}$ with only (partial) optimisation over sky location, fixing the angular size to a fiducial value. In the associated cosmological analyses, these tSZ observables are modelled differently. All three sets of analyses assume that $q_{\mathrm{opt}}$ follows a unit-variance Gaussian distribution. The mean it is assigned, however, differs from analysis to analysis. The SPT cluster number counts analyses \citep{Vanderlinde2010,Bleem2015,Bocquet2018} account for the optimisation bias at the likelihood level, using as the expected value of $q_{\mathrm{opt}}$ (their `biased significance', $\xi$) $\bar{q}_{\mathrm{opt}} = (\bar{q}_{\mathrm{t}}^2 + 3 )^{1/2}$. This corresponds to our formula for $\bar{q}_{\mathrm{opt}}$, given in Eq. (\ref{correction}), if we set $\Delta \sigma^2 = 0$ and $f_{\mathrm{eff}} = f$; a rather heuristic justification for this prescription is given in \citet{Vanderlinde2010}. The \textit{Planck} cluster number counts analyses \citep{Planck2014XX,Ade2016}, despite using $q_{\mathrm{opt}}$ as an observable, do not account for the optimisation bias at all, as neither does the reanalysis with CMB lensing mass calibration of \citet{Zubeldia2019}. Finally, the ACT analysis \citep{Hasselfield2013}, which uses $q_{\mathrm{opt}}$ at a fixed angular scale in order to mitigate against the optimisation bias, does not account for optimisation over sky location, which means that in principle it suffers from the bias, although from a somewhat smaller one.

Thanks to the insights gained in Section \ref{sec:theory}, it should be clear that if the signal-to-noise is maximised over some set of parameters, the consequences of doing so (non-Gaussianity, standard deviation different from unity, and optimisation bias) should be carefully quantified and, if deemed non-negligible, properly accounted for. As an illustration of how this can be done, as well as of the general theoretical arguments presented in Section \ref{sec:theory}, in the next section we quantify these effects for an idealised SZ experiment using a set of MC mock observations.

\subsection{Quantifying the statistics of $q_{\mathrm{opt}}$ with MC observations}\label{mc_test}

\subsubsection{Production of MC data}\label{sec:dataproduction}

In this section, we quantify the statistics of $q_{\mathrm{opt}}$ for a set of idealised tSZ observations produced with MC simulations. As our signal we take the Compton-$y$ map due to a set of galaxy clusters of some given masses $M_{500}$ (defined, as customary, as the mass within $R_{500}$, the radius within which the mean enclosed density is $500$ times the critical density at the cluster's redshift) and at some redshift $z$. We take our clusters to be at $z=0.3$ and consider a set of 16 masses such that the mean signal-to-noise $\bar{q}_{\mathrm{t}}$, as given by Eq. (\ref{snr}) after ensemble averaging over noise realisations, is logarithmically spaced between $\bar{q}_{\mathrm{t}} = 3$ and $\bar{q}_{\mathrm{t}} = 30$ (see our assumed experimental specifications below). We assume that our clusters have the pressure profile of \citet{Arnaud2010} (used, e.g., in the \textit{Planck} analyses), from which the Compton-$y$ signal at a given sky location can be computed,

\begin{equation}\label{compton}
     y = \frac{\sigma_{\mathrm{T}}}{m_{\mathrm{e}}c^2} \int P dl,
\end{equation}
where $P$ is the cluster electron pressure, $\sigma_{\mathrm{T}}$ is the electron Thomson scattering cross-section, $m_{\mathrm{e}}$ is the electron mass, and $dl$ is the proper distance element along the line of sight. Following the \textit{Planck} analyses, we fix the concentration to $c = 1.177$ (e.g., \citealt{Planck2014XX}).

Adopting this cluster model, our signal vector $s_0 \bmath{s}$ is the cluster Compton-$y$ parameter map, as given by Eq. (\ref{compton}), convolved with a Gaussian beam with a FWHM of $5$\,arcmin and evaluated on a square map of $128 \times 128$ pixels. We use a pixel size of $1.7$\,arcmin and place the cluster at the centre of the map. We then add Gaussian white noise with a standard deviation per pixel of $\sigma_{\mathrm{pixel}} = 8.83 \times 10^{-7}$, which corresponds to a noise level of $\sigma_n = 1.5 \times 10^{-6}$\,arcmin. That is, our noise covariance matrix $\mathbfss{C}$ is diagonal, with all diagonal elements equal to $\sigma_{\mathrm{pixel}}^2$. This procedure generates a simulated data vector $\bmath{d}$ within the formalism of Section \ref{sec:theory}. We generate $10^5$ such mock data vectors for each of our $16$ values of $\bar{q}_{\mathrm{t}}$.

We then use our signal, $s_0 \bmath{s}$, in order to construct a matched filter template, $\bmath{s}$, using as the amplitude (or normalisation) parameter $s_0$ the value of the signal at $R_{500}$. We regard this template as a function of three parameters $\bmath{\theta}$: the cluster angular size, $\theta_{500}$, defined as $R_{500}/d_A$, where $d_A$ is the angular diameter distance to the cluster, and the two angular coordinates of its centre with respect to the centre of the map (i.e., with respect to the true cluster centre), $\theta_x$ and $\theta_y$. That is, within the theoretical framework of Section \ref{sec:theory}, the number of additional fitting parameters (or `template parameters') is $f=3$. Throughout this section we assume the noise covariance to be known; in Section \ref{sec:cov}, however, we consider the impact of noise covariance estimation. Using the matched filter constructed this way, for each of our mock data vectors we make a signal-to-noise measurement at the true parameter values ($\theta_{500} =  \theta_{500}^{\mathrm{true}}$ and $\theta_x = \theta_y = 0$), $q_{\mathrm{t}}$, an optimal signal-to-noise $q_{\mathrm{opt}}$, obtained by maximising the signal-to-noise over the three template parameters, and a partially-optimal signal-to-noise $q_{\mathrm{p-opt}}$, obtained by evaluating the signal-to-noise on a parameter grid and picking the maximum value. For this last set of measurements we use the parameter grid spanned by 80 values of $\theta_{500}$ logarithmically spaced between  $\theta_{500} = 0.94$\,arcmin and $\theta_{500} = 101.57$\,arcmin and the angular coordinates of the centres of each pixel. This procedure generates a set of $q_{\mathrm{t}}$, $q_{\mathrm{opt}}$ and $q_{\mathrm{p-opt}}$ mock measurements generated according to the formalism of Section \ref{sec:theory}. For each data vector we also compute $-2 \ln \mathcal{L}$ for four different cases: (i) at the true value of the template parameters $\bmath{\theta}$ \textit{and} of the amplitude parameter $s_0$, i.e., for the true model (the `true case'), (ii) at the true value of $\bmath{\theta}$ and the matched filter estimate of $s_0$ at that value of $\bmath{\theta}$, $\hat{s}_0 (\bmath{\theta}^{\mathrm{true}})$ (the `fixed case'), (iii) at the matched filter solution for $\bmath{\theta}$, $\hat{\bmath{\theta}}$, and its corresponding matched filter estimate of $s_0$, $\hat{s}_0 (\hat{\bmath{\theta}})$ (the `optimal case'), and (iv), at the partially-optimised parameter values and the corresponding matched filter estimate (the `partially optimal' case).

We note that our setup may not seem equivalent to what is usually done in practice in order to detect and characterise galaxy clusters in mm surveys, which is to use multi-frequency matched filters (MMF) that operate directly on the experiment frequency maps, rather than on an estimated Compton-$y$ map (e.g., \citealt{Planck2014XX}). In Appendix \ref{appendix}, however, we show that if the Compton-$y$ map is obtained through an Internal Linear Combination (ILC; see, e.g., \citealt{Delabrouille2007}), these two approaches are mathematically equivalent.

\subsubsection{Standard deviation and non-Gaussianity}

\begin{figure}
\centering
\includegraphics[width=0.5\textwidth]{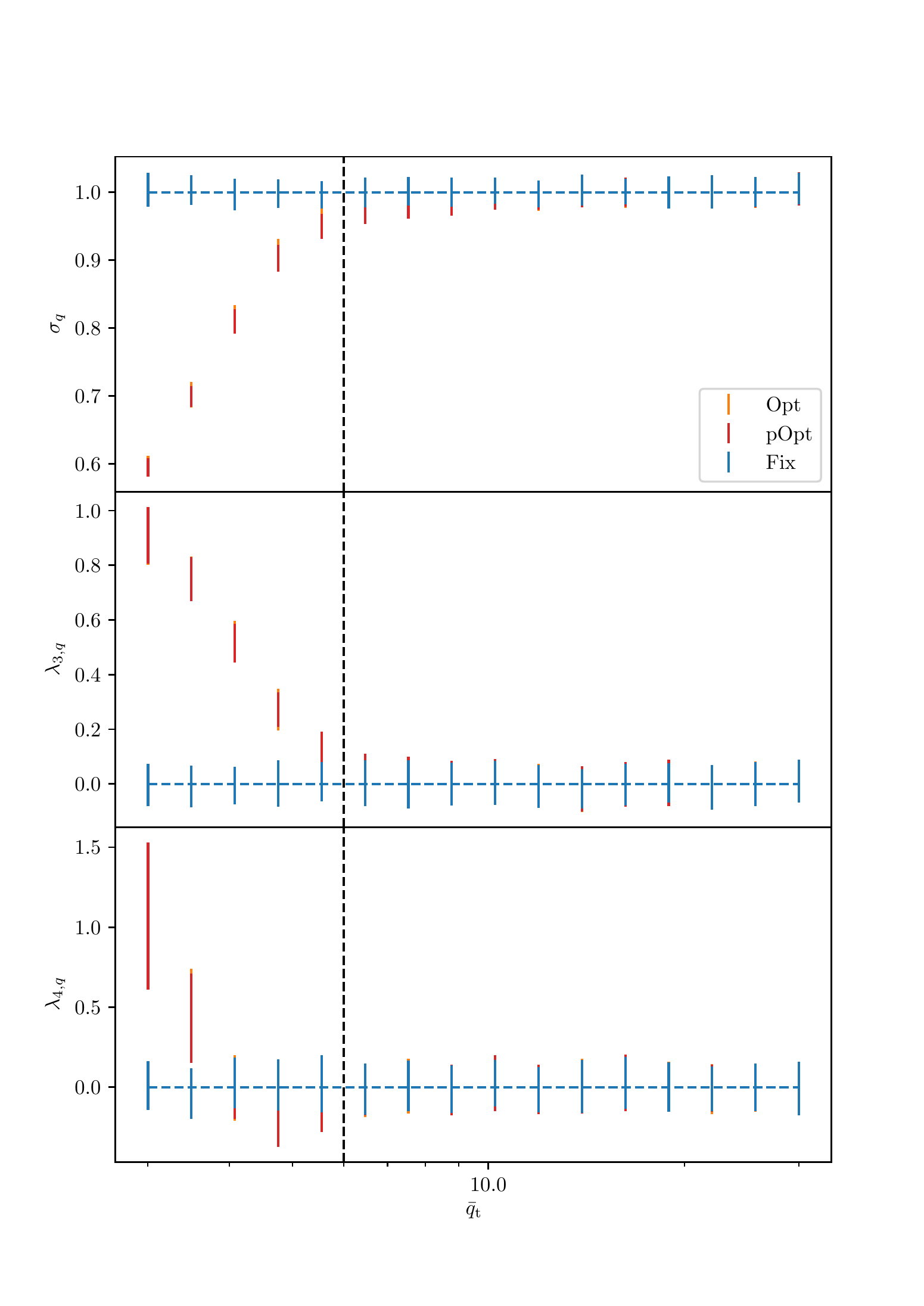}
\caption{Standard deviation (top panel), reduced skewness (middle panel), and reduced kurtosis (bottom panel) of $q_{\mathrm{opt}}$ (orange error bars), $q_{\mathrm{p-opt}}$ (red error bars), and $q_{\mathrm{t}}$ (blue error bars) as a function of true mean signal-to-noise, $\bar{q}_{\mathrm{t}}$, as estimated with our set of mock observations. The reference values of $\sigma_q = 1$, $\lambda_{3,q} = 0$, $\lambda_{4,q} = 0$, and $\bar{q}_{\mathrm{t}} = 6$ are shown as straight lines. We note that the error bars have been increased by a factor of 10 for better visualisation. }
\label{fig:statistics}
\end{figure}

We first investigate the standard deviation of $q_{\mathrm{opt}}$ and $q_{\mathrm{p-opt}}$ and the extent of their departure from Gaussianity. Fig. \ref{fig:statistics} shows the standard deviation $\sigma_q$, reduced skewness $\lambda_{3,q}$, and reduced kurtosis $\lambda_{4,q}$ of the distribution of $q_{\mathrm{opt}}$ (orange error bars) and of $q_{\mathrm{p-opt}}$ (red error bars) as a function of true mean signal-to-noise $\bar{q}_{\mathrm{t}}$, as estimated with our set of mock measurements. We note that the error bars of the estimates of the standard deviation have been increased by a factor of 10 to allow for better visualisation. It can be seen that these three statistics are essentially the same for both $q_{\mathrm{opt}}$ and $q_{\mathrm{p-opt}}$. Both $q_{\mathrm{opt}}$ and $q_{\mathrm{p-opt}}$ become progressively more Gaussian as $\bar{q}_{\mathrm{t}}$ is increased, their skewness and kurtosis being driven to zero. In addition, their standard deviation approaches unity with increasing $\bar{q}_{\mathrm{t}}$. Roughly, for values of $\bar{q}_{\mathrm{t}}$ greater than $\simeq 6$, $q_{\mathrm{opt}}$ can be approximately taken to follow a unit-variance Gaussian distribution. For comparison, the standard deviation, skewness, and kurtosis of $q_{\mathrm{t}}$, as estimated from our mock measurements, are also shown. As expected, the standard deviation is consistent with unity throughout, as are the skewness and kurtosis with zero.

\subsubsection{Effective number of fitting parameters and optimisation bias}\label{sec:biasplot}

\begin{figure}
\centering
\includegraphics[width=0.5\textwidth]{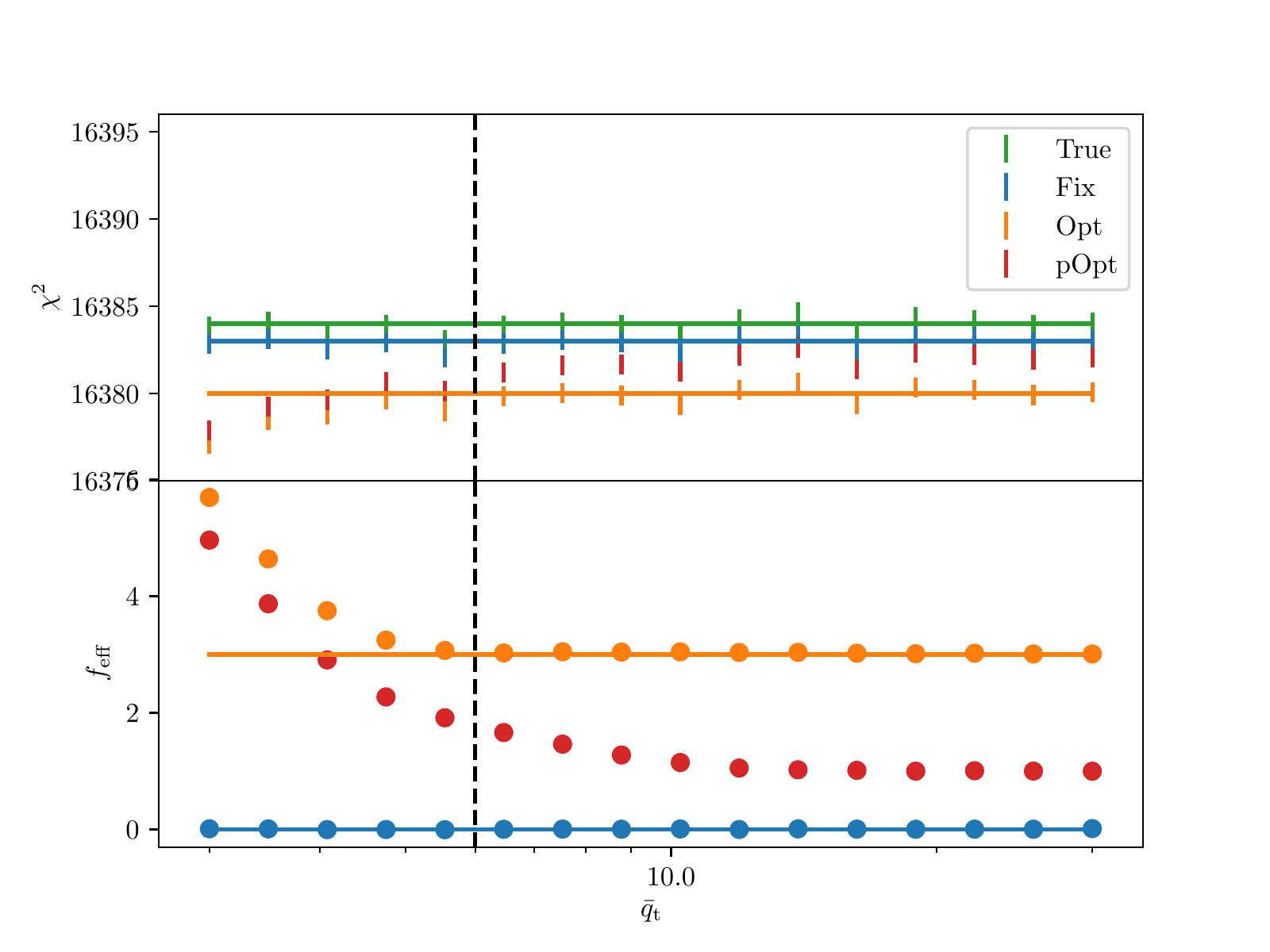}
\caption{\textit{Top panel}: Expected value of $-2 \ln \mathcal{L}$ for the true, fixed, optimal, and partially optimal cases (green, blue, orange, and red error bars, respectively) as a function of the mean true signal-to-noise, $\bar{q}_{\mathrm{t}}$, as estimated with our mock observations, shown along their associated reference values of $16384$, $16383$, $16382$, and $16380$, respectively. \textit{Bottom panel}: Effective number of additional fitting parameters, $f_{\mathrm{eff}}$, for the fixed, optimal, and partially optimal cases (blue, orange, and red error bars, respectively) as a function of $\bar{q}_{\mathrm{t}}$, as estimated with our mock observations, shown along their reference values of $0$, $1$ and $3$. The reference value of $\bar{q}_{\mathrm{t}} = 6$ is also shown.}
\label{fig:dof}
\end{figure}

We now quantify, as a function of the mean signal-to-noise, the effective number of additional fitting parameters $f_{\mathrm{eff}}$ associated with optimising the signal-to-noise over our three template parameters. We do so with our measured values of $-2 \ln \mathcal{L}$. The top panel of Fig.\,\ref{fig:dof} shows the empirical mean of this quantity as a function of $\bar{q}_{\mathrm{t}}$ for the four cases for which we have computed it: the `true case' (green error bars), the `fixed case' (blue error bars), the optimal case (orange error bars), and the partially optimal case (red error bars). For the true case, for which $-2 \ln \mathcal{L}$ is $\chi^2$-distributed with $n$ degrees of freedom, where $n$ is the number of data points, the estimated mean is consistent with $16384 = 128^2$ throughout, as expected. Similarly, for the fixed case, for which $-2 \ln \mathcal{L}$ is also $\chi^2$-distributed, though with $n-1$ degrees of freedom, the estimated mean is consistent with $128^2 - 1$, also as expected. However, for the optimal and partially optimal cases, the situation is different, as a result of $-2 \ln \mathcal{L}$ no longer being $\chi^2$-distributed. For the optimal case, the estimated mean of $-2 \ln \mathcal{L}$ is consistent with $128^2 - 4 = 16380$ for large values of $\bar{q}_{\mathrm{t}}$ (roughly, for $\bar{q}_{\mathrm{t}} > 6$), as it would be naively (and wrongly) expected for three additional fitting parameters, but it drops to lower values for small values of $\bar{q}_{\mathrm{t}}$: at the low signal-to-noise end, our model `fits the data better' than a linear model would. For the partially optimal case, the estimated mean of $-2 \ln \mathcal{L}$ is consistent with $128^2 - 2 = 16380$ for large values of $\bar{q}_{\mathrm{t}}$, dropping to lower values for smaller values of $\bar{q}_{\mathrm{t}}$, as in the optimal case. The high signal-to-noise behaviour, indicative of just one additional fitting parameter, can be explained by the fact that, at high signal-to-noise, the true central pixel is almost always chosen in the grid optimisation process, and therefore no overfitting arising from location can take place. Effectively, there is just one additional fitting parameter, the cluster angular size. We note that this is a feature of our true signal being placed in the centre of a pixel, which allows for the possibility of the grid optimisation algorithm matching the sky location perfectly. This will obviously not be the case in a real experimental setting, in which there will always be an offset between the centre of the chosen pixel and the true cluster sky location. This implies that, in a real setting, the fit will be worse, and $-2 \ln \mathcal{L}$ will not converge to $n^2-2$, but will take larger values. We note that true optimisation does not suffer from this artefact, which means that our results in that instance are applicable to a real experimental setting in which the cluster may not be centred in the centre of a pixel.

In the bottom panel of Fig. \ref{fig:dof} we rephrase the problem in terms of the effective number of additional fitting parameters, $f_{\mathrm{eff}}$, where we recall that $\left\langle -2 \ln \mathcal{L} \right\rangle = n-1-f_{\mathrm{eff}}$. As expected from the top panel, for the fixed case, $f_{\mathrm{eff}}$ is consistent with zero throughout; for the optimal case, $f_{\mathrm{eff}} \simeq 3$ for large values of $\bar{q}_{\mathrm{t}}$ (roughly, for $\bar{q}_{\mathrm{t}} > 6$), progressively taking larger values for smaller values of $\bar{q}_{\mathrm{t}}$; and for the partially optimal case, $f_{\mathrm{eff}} \simeq 1$ for large values of $\bar{q}_{\mathrm{t}}$, similarly taking larger values with decreasing $\bar{q}_{\mathrm{t}}$.

Now that we have determined the standard deviation of $q_{\mathrm{opt}}$ and $f_{\mathrm{eff}}$ as a function of $\bar{q}_{\mathrm{t}}$, we are in a position to compute $\bar{q}_{\mathrm{opt}}$ using Eq. (\ref{correction}). We do so by interpolating over our values of $\sigma_q$ and $f_{\mathrm{eff}}$. Fig. \ref{fig:prediction} shows such prediction (shown as $\bar{q}_{\mathrm{opt}}/\bar{q}_{\mathrm{t}}$) as a function of $\bar{q}_{\mathrm{t}}$ (solid orange curve), along with the associated direct estimates of $\bar{q}_{\mathrm{opt}}$ obtained by ensemble averaging over our $q_{\mathrm{opt}}$ measurements (orange error bars). To illustrate the relative importance of accounting for the deviation form unity in the variance and for the change in $f_{\mathrm{eff}}$ from the `naive' value of $f = 3$, we also show the curve that is obtained if we ignore the change in the variance and set $f_{\mathrm{eff}} = f = 3$, which we refer to as the $f$ correction (dotted orange curve), and the one that is obtained if we ignore the change in the variance but use the MC-estimated value of $f_{\mathrm{eff}}$ (dashed orange curve). It can be seen that for, roughly, $\bar{q}_{\mathrm{opt}} > 6$, $\bar{q}_{\mathrm{opt}}$ is essentially given by $\bar{q}_{\mathrm{opt}} \simeq (\bar{q}_{\mathrm{t}}^2 + f)^{1/2}$, i.e., by the $f$ correction alone. This was to be expected from our MC measurements of $\sigma_q$ and of $f_{\mathrm{eff}}$, which we recall converge to 1 and to $f$, respectively, by $\bar{q}_{\mathrm{opt}} \simeq 6$. This constitutes a remarkable result, as it means that for signal-to-noise values similar to those typically used as selection thresholds in cosmological tSZ catalogues (e.g., in \textit{Planck}, a threshold of $q_{\mathrm{opt}}=6$ was chosen), as well as for higher values, $\bar{q}_{\mathrm{opt}}$ can be computed accurately (e.g., in a likelihood) in a straightforward way without the need to resort to MC data. If a more accurate expression at low values of $\bar{q}_{\mathrm{opt}}$ is needed (e.g., if a lower selection threshold is used in an analysis), one could interpolate over the numerical results, either over $f_{\mathrm{eff}}$ and $\sigma_q$, or directly over $\bar{q}_{\mathrm{opt}}$; alternatively, one could fit a function (e.g., a polynomial) to them.

We remark that we have obtained our numerical results for the case in which three parameters (one angular size parameter and two sky coordinates) are optimised over. This was the case of the \textit{Planck} and SPT analyses (e.g., \citealt{Planck2016xxvii,Bocquet2018}), and we thus expect our results to be relevant for them. In ACT, on the other hand, the signal-to-noise over was optimised only over two sky coordinates (e.g, \citealt{Hilton2020}). Since this involves less parameters than the case considered here, and since these are the same as two of our three parameters, we expect our results to also be relevant for ACT. In particular, the $f$ correction should be accurate in a similar, if not broader, signal-to-noise range, noting that, in this case, $f=2$.

\subsubsection{Gridding bias}\label{sec:gridbias}

\begin{figure*}
\centering
\includegraphics[width=0.8\textwidth]{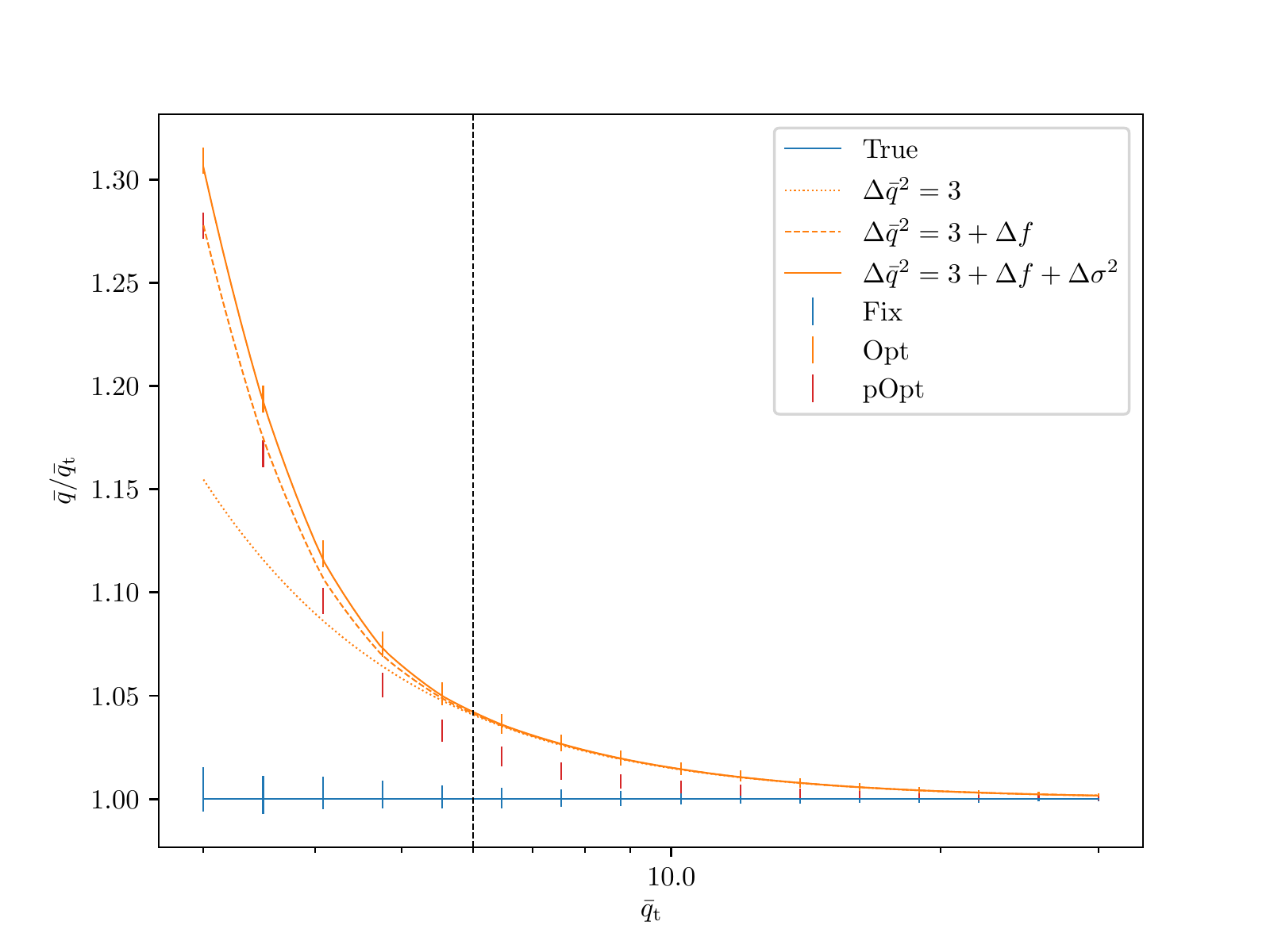}
\caption{Empirical means of our $q_{\mathrm{opt}}$, $q_{\mathrm{p-opt}}$, and $q_{\mathrm{t}}$ measurements (orange, red, and blue error bars, respectively), shown along with three predictions for the mean of $q_{\mathrm{opt}}$, $\bar{q}_{\mathrm{opt}}$: the full prediction given by Eq. \ref{correction} (solid orange curve), that given by Eq. \ref{correction} neglecting the variance change term (i.e., $\Delta \sigma^2=0$) (dashed orange curve), and that given by the simple $f$ correction, with $f_{\mathrm{eff}} = f =3$ (dotted orange curve). All the error bars have been multiplied by a factor of 10 to allow for better visualisation, and the reference value of $\bar{q}_{\mathrm{t}} = 6$ is shown as a vertical dashed line. As noted in Section \ref{sec:biasplot}, the $f$ correction is the most significant contribution to $\bar{q}_{\mathrm{opt}}$ for $\bar{q}_{\mathrm{t}} \gtrsim 6$.}
\label{fig:prediction}
\end{figure*}

Fig. \ref{fig:prediction} also shows the empirical means of our $q_{\mathrm{p-opt}}$ and $q_{\mathrm{t}}$ measurements as a function of $\bar{q}_{\mathrm{t}}$, shown, too, normalised by $\bar{q}_{\mathrm{t}}$ (red and blue error bars, respectively). As expected, the mean of the $q_{\mathrm{t}}$ measurements is consistent with $\bar{q}_{\mathrm{t}}$ and  the mean of $q_{\mathrm{p-opt}}$ is always below that of $q_{\mathrm{opt}}$, as, due to gridding, there is less `overfitting'. This can be understood as a `gridding bias' acting in the direction opposite to the optimisation bias (in the context of just sky location, this would be the usual miscentering bias). We remark that this bias depends on the grid chosen: a coarser grid would lead to a larger gridding bias, which could even lead to an overall negative bias, whereas a grid with infinitesimal cells would yield a vanishingly small gridding bias, $q_{\mathrm{p-opt}}$ being equal to $q_{\mathrm{opt}}$. Due to the presence of this additional bias, which does not exist for true optimisation, and to the fact that for true optimisation the optimisation bias can be accurately modelled with the simple $f$ correction for reasonably large values of signal-to-noise, we advocate for the use of the `true' optimal signal-to-noise, $q_{\mathrm{opt}}$, as a cluster observable, in detriment to the commonly-used partially-optimal signal-to-noise. In practice, this can be realised, if not with an optimisation algorithm, with the use of an appropriately finely-spaced grid.

Regarding the spatial gridding, we should note that in our mock observations, each pixel receives an independent noise realisation; as a consequence, the `optimal limit' is not reached by evaluating the signal-to-noise at the centre of each pixel, as there is structure in the data down to the pixel level. This is not typically the case in real maps, which tend to be well oversampled: the `optimal limit' in the angular coordinates may therefore be reached by just evaluating the signal-to-noise at the centre of each pixel. Care should still be taken, however, with the grid optimisation of the angular scale parameter (and of any other additional parameters).

 \subsubsection{Parameter estimates: $\theta_{500}$ and sky position}\label{sec:paramest}

\begin{figure}
\centering
\includegraphics[width=0.5\textwidth]{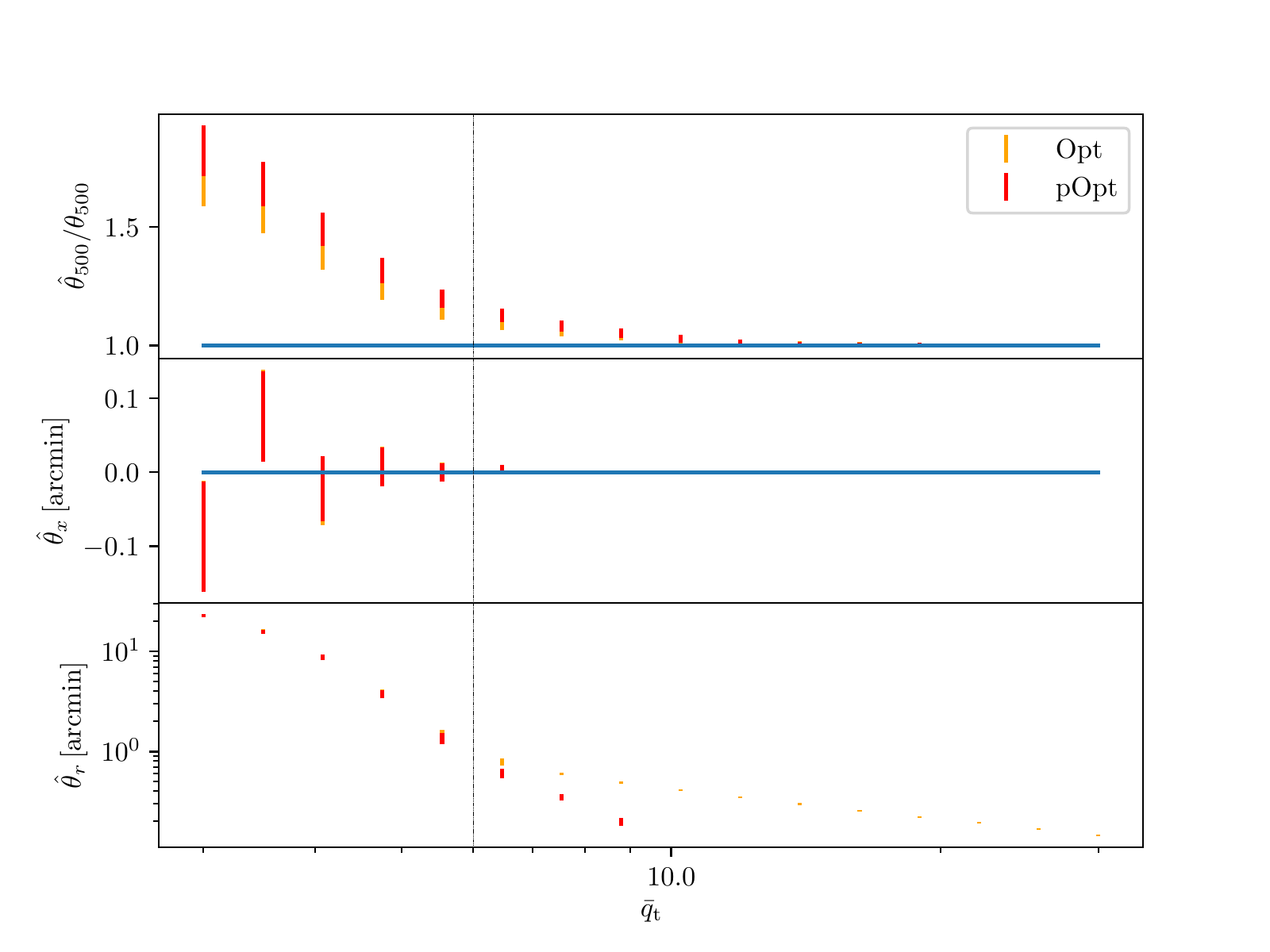}
\caption{Matched filter estimates of $\bmath{\theta}_{500}$, $\theta_x$, and $\theta_r$, as obtained from our mock data for both true and partial optimisation (orange and red error bars, respectively).}
\label{fig:parameters}
\end{figure}

 Although the signal-to-noise is usually the main quantity of interest in tSZ galaxy cluster studies, maximising the matched filter signal-to-noise over a set of parameters $\bmath{\theta}$ also yields an estimator for $\bmath{\theta}$, $\hat{\bmath{\theta}}$, which we refer to as the matched filter solution. As shown in Section \ref{sec:theory}, this matched filter solution is equal to the maximum-likelihood estimator for $\bmath{\theta}$, and therefore it is, in general, biased (see Section \ref{sec:solutionbias}). As an illustration of this point, Fig. \ref{fig:parameters} shows the empirical means of the matched filter solutions of two of our parameters, $\theta_{500}$ (upper panel) and $\theta_x$ (middle panel), as a function of $\bar{q}_{\mathrm{t}}$, as obtained with our mock measurements for true optimisation (orange error bars) and partial optimisation (red error bars). (The results for $\theta_y$ are analogous to those of $\theta_x$ and hence not shown.) It can be observed that the matched filter estimates of $\theta_{500}$ are biased high for both true and partial optimisation, the bias decreasing with increasing signal-to-noise. The estimates of $\theta_x$, on the other hand, are not biased, as expected from the symmetry of the problem.

 In order to illustrate the point made in Section \ref{sec:solutionbias} that the extent, and even the existence, of a bias in the parameter estimates is entirely parametrisation-dependent, the bottom panel of Fig. \ref{fig:parameters} shows the empirical means of the parameter $\theta_r \equiv (\theta_x^2 + \theta_y^2)^{1/2}$, which along with $\phi \equiv \tan^{-1} (\theta_y/\theta_x)$ constitute a fully equivalent parametrisation to $\theta_x$ and $\theta_y$. As expected, the estimates of $\theta_r$ are biased high, as they receives contributions from the second-order moments of $\theta_x$ and $\theta_y$. We note that for the partially-optimised case, the bias disappears at high signal-to-noise. This is an artefact of our simulated clusters being centred at the centre of a pixel, which allows, at the high signal-to-noise end, for the true location to be chosen perfectly by the grid optimisation algorithm. This will of course not happen in a real experimental setting (see Section \ref{sec:biasplot} for a similar discussion).

 \subsubsection{Dependence on redshift and on noise levels}

 The results presented in the previous sections were obtained as a function of mean true signal-to-noise, $\bar{q}_{\mathrm{t}}$, by fixing the cluster redshift at $z=0.3$ and varying the cluster mass $M_{500}$. There is, of course, a perfect degeneracy in $\bar{q}_{\mathrm{t}}$, whereby a given value of $\bar{q}_{\mathrm{t}}$ can be obtained by moving along some curve in the $M_{500}$--$z$ plane. However, although $\bar{q}_{\mathrm{t}}$ may be perfectly degenerate along this curve, the actual observation (the Compton-$y$ map) will not be so. As a consequence, the statistics of $q_{\mathrm{opt}}$ and, in particular, the magnitude of the optimisation bias are not guaranteed to depend exclusively on $\bar{q}_{\mathrm{t}}$, but should in general be regarded as a function of both $M_{500}$ and $z$. One can also regard them as dependent on the assumed observational specifications, e.g., the white noise levels, under which $\bar{q}_{\mathrm{t}}$ is also perfectly degenerate, with the data not being so either. This poses the question of to what extent can our results, especially the magnitude of the optimisation bias, be regarded as a function of only $\bar{q}_{\mathrm{t}}$, i.e., whether the higher-dimensional dependency on $M_{500}$, $z$, noise levels, etc. can be safely condensed into a dependency on just $\bar{q}_{\mathrm{t}}$.

 In order to address this question, we produce a new set of mock observations following exactly the same procedure as the one detailed in Section \ref{sec:dataproduction}, but for a set of 16 points lying in the curve of the $M_{500}$--$z$ plane defined by $\bar{q}_{\mathrm{t}} (M_{500},z) = 6$. Specifically, we consider 16 values of $z$ linearly spaced between $z = 0.1$ and $z = 0.8$ and compute the corresponding value of $M_{500}$ that yields $\bar{q}_{\mathrm{t}} = 6$. We produce $10^4$ mock data vectors for each point along this curve, and we do so for three different white noise levels: $\sigma_n = 1.5 \times 10^{-6}$\,arcmin (the same as used for the mock data of the previous sections), $\sigma_n = 3 \times 10^{-6}$\,arcmin, and $\sigma_n = 0.75 \times 10^{-6}$\,arcmin. Here we consider only true optimisation, computing the value of $q_{\mathrm{opt}}$ for each mock data vector.

\begin{figure}
\centering
\includegraphics[width=0.5\textwidth]{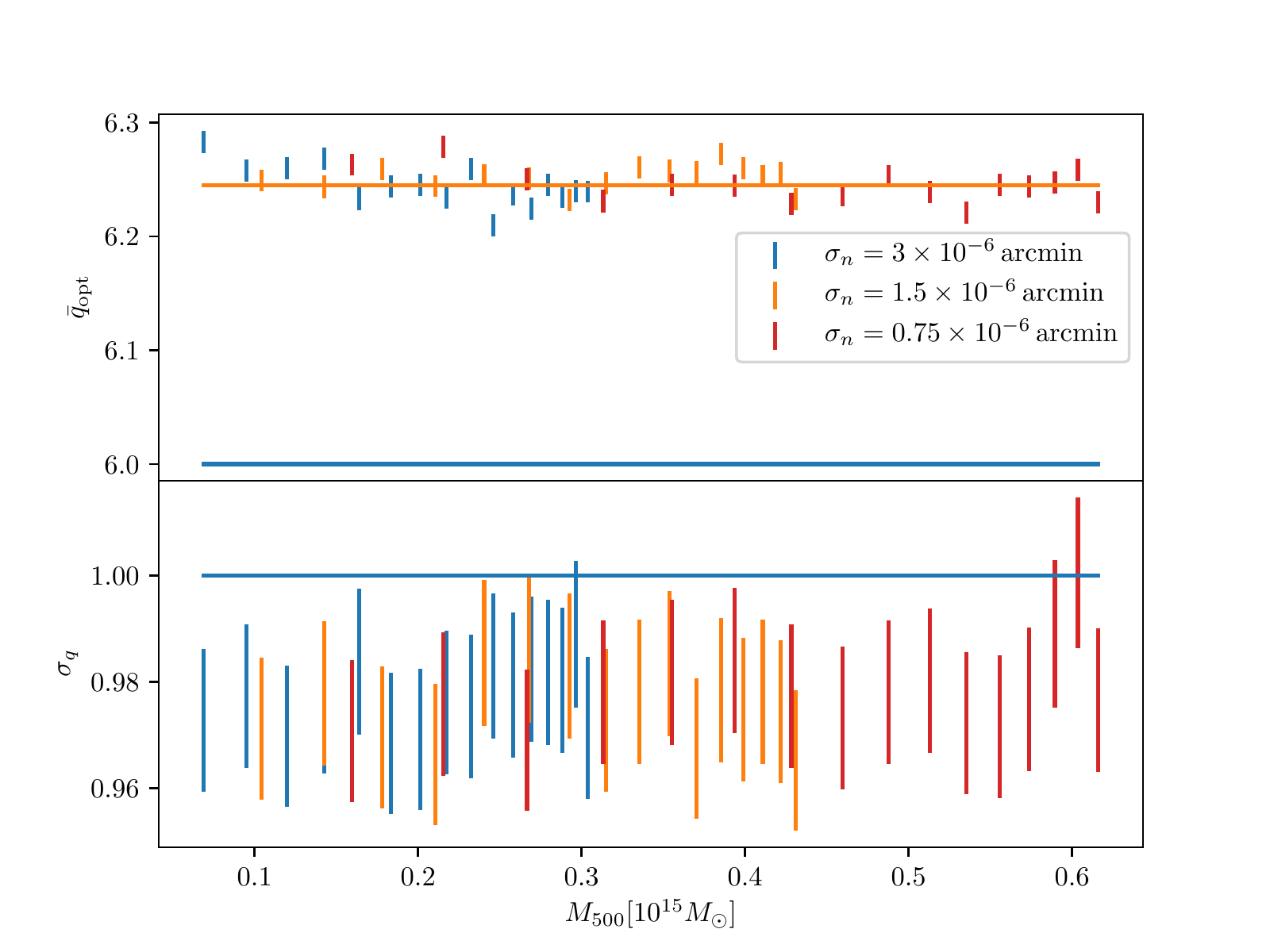}
\caption{Empirical mean (upper panel) and standard deviation (lower panel) of our $q_{\mathrm{opt}}$ measurements produced along the $\bar{q}_{\mathrm{t}} (M_{500},z) = 6$ curve for three different noise levels (blue, orange, and red error bars). In the upper panel, the reference values of  $q = 6$ and $q= (6^2 + 3)^{1/2}$ are shown as straight lines (blue and orange lines, respectively), as is $\sigma_q = 1$ in the lower panel (blue line).}
\label{fig:6}
\end{figure}

 Fig. \ref{fig:6} shows the empirical mean (upper panel) and standard deviation (lower panel) of these $q_{\mathrm{opt}}$ measurements as a function of $M_{500}$ for the the three instances of noise levels (blue, orange, and red error bars). The upper panel also shows as straight lines the reference values of $q = 6$, which is the value of $\bar{q}_{\mathrm{t}}$ for all the measurements (blue line), and that of $\bar{q}_{\mathrm{opt}}$ computed with only the $f$ correction, i.e., $\bar{q}_{\mathrm{opt}} = (\bar{q}_{\mathrm{t}}^2 + f)^{1/2}$ with $f=3$, which we know is an accurate approximation for $\bar{q}_{\mathrm{opt}}$ at $\bar{q}_{\mathrm{t}} = 6$ (orange line). On the other hand, the lower panel also shows the reference value of unity standard deviation (blue line). It can be seen that any variation of $\bar{q}_{\mathrm{opt}}$ and $\sigma_q$ across the curve $\bar{q}_{\mathrm{t}} (M_{500},z) = 6$ is clearly subdominant with respect to the overall differences with respect to $\bar{q}_{\mathrm{t}}=6$ and $\sigma_q = 1$, respectively, and that this is true for the three noise levels that we have considered. We expect this to also be the case for $\bar{q}_{\mathrm{t}} > 6$, as with increasing $\bar{q}_{\mathrm{t}}$ the $f$ correction becomes progressively more accurate  and $\sigma_q$ gets closer to unity. In brief, within the mass--redshift and noise levels range that we have considered, the optimisation bias and $\sigma_q$ can be safely regarded as just a function of  $\bar{q}_{\mathrm{t}}$ for, at least, $\bar{q}_{\mathrm{t}} > 6$. 

 \subsection{Noise covariance estimation and other biases}\label{sec:cov}

\begin{figure}
\centering
\includegraphics[width=0.5\textwidth]{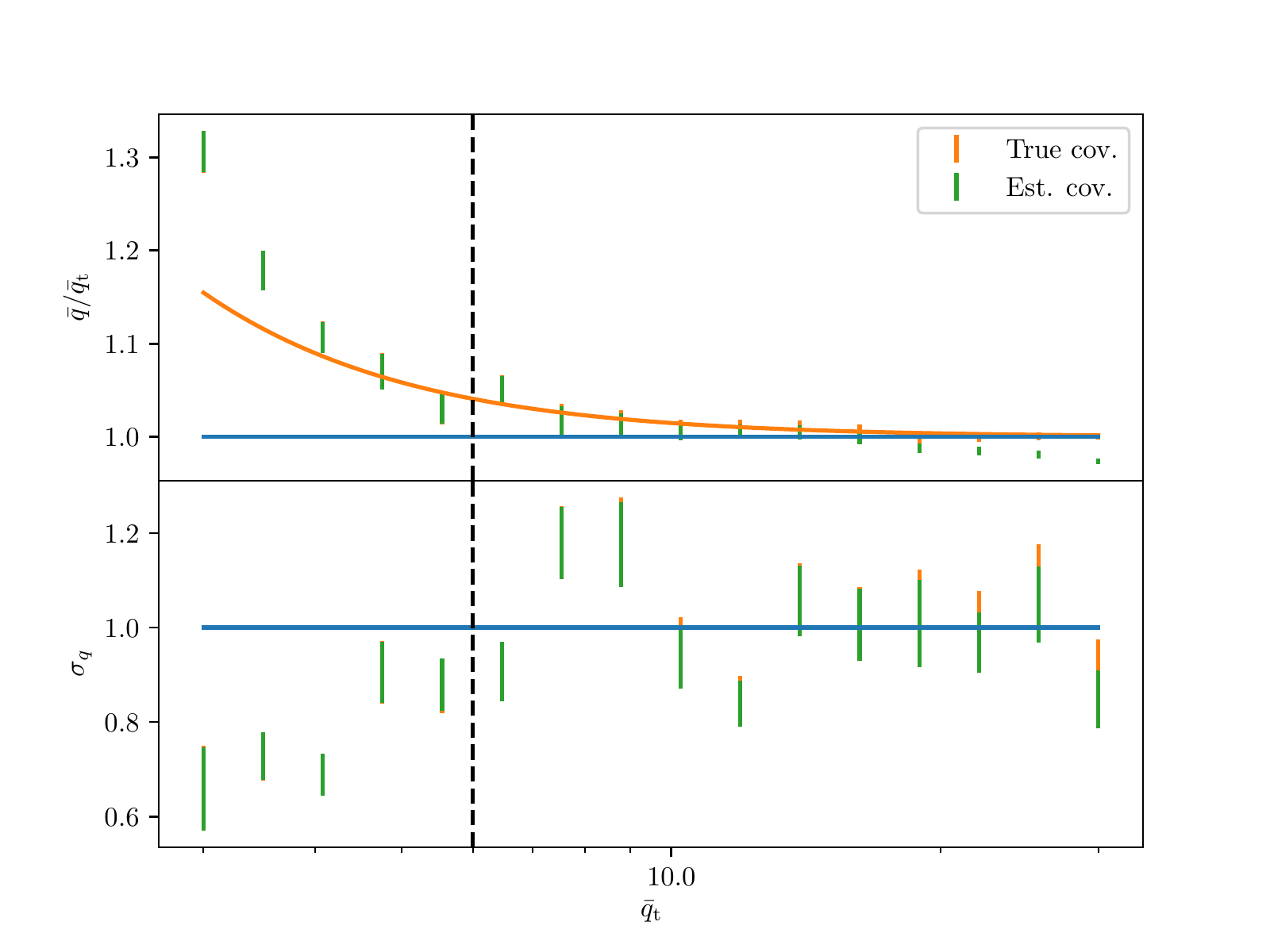}
\caption{Empirical mean and standard deviation of our $q_{\mathrm{opt}}$ measurements produced using the true noise covariance (orange error bars) and estimating it from the data (green error bars).}
\label{fig:covariance}
\end{figure}

 Throughout this paper we have assumed the noise covariance to be perfectly known. In a real setting, however, the noise covariance is typically estimated from the data itself. If the signal to be extracted is small enough, it can be taken to be equal to the total data covariance. This was the case, e.g., of the \textit{Planck} analysis, in which the noise covariance was taken to be the empirically-estimated data covariance \citep{Planck2016xxvii}. While this is a good approximation at the low signal-to-noise end, it is no longer so for larger signal-to-noise values, where the signal becomes comparable to the noise, potentially leading to a bias in the retrieved signal. This is illustrated in Figure \ref{fig:covariance}, which shows the empirical mean and standard deviation of two sets of $q_{\mathrm{opt}}$ mock measurements as a function of $\bar{q}_{\mathrm{t}}$ (1000 data points for each value of $\bar{q}_{\mathrm{t}}$ each). In order to generate these two data sets, we produced mock data vectors (Compton-$y$ maps) with the same specifications as used in Section \ref{mc_test} and we then applied a matched filter on each of them, maximising the signal-to-noise over angular size and sky location. For one of the data sets (orange error bars), the noise covariance used to construct the matched filter was the true noise covariance (i.e., this mock data set is equivalent to that analysed throughout Section \ref{mc_test}). For the other data set (green error bars), the noise covariance was taken to be equal to the data covariance, which was estimated empirically from the data vector by taking its standard deviation (that is, assuming the noise to be white). Is is apparent that while the empirical means of these two sets are essentially the same at low signal-to-noise, where noise is dominant, they deviate at the high signal-to-noise end, with noise estimation biasing $q_{\mathrm{opt}}$ low. This negative bias has a simple explanation (see, e.g.,  Eq. \ref{snr}): as the noise covariance is overestimated, the signal-to-noise is underestimated. The standard deviation is also affected at the high signal-to-noise end, although less so. Approaches in which an estimate of the signal is subtracted from the data, allowing for better noise covariance estimation (something that can be done, e.g., iteratively), can be followed to mitigate against this bias. An assessment of these approaches is, however, beyond the scope of this paper.

Other potential sources of bias include, but are not limited to, the wrong modelling of the signal to be extracted, i.e., the matched filter not matching the true cluster tSZ signal, both spatially and spectrally, the latter possible if the relativistic SZ effect (e.g., \citealt{Sazonov1998, Challinor1998, Chluba2012SZpack}) is non-negligible, and the presence of other, `foreground' signals such as the CMB, the kSZ effect, radio emission from the cluster's galaxies, and point sources (see, e.g., \citealt{Mroczkowski2019,Erler2019}). Any assessment of these potential sources of systematics is beyond the scope of this paper.

\subsection{Impact of optimisation on cluster number counts: \textit{Planck} and the Simons Observatory}\label{planck}

We now shift our attention to the impact of the effects of optimisation that were described theoretically in Section \ref{sec:theory} and illustrated numerically in Section \ref{mc_test} may have on cluster number counts analyses that use $q_{\mathrm{opt}}$ as an observable. We will focus solely on the optimisation bias, as indeed the non-Gaussianity of $q_{\mathrm{opt}}$ and the deviation of its standard deviation from unity are small for signal-to-noise values of about $6$ and above, which is the typical selection threshold in such analyses. In particular, we will determine whether the optimisation bias can be safely neglected in the analysis of cluster samples from \textit{Planck} and the Simons Observatory (SO; \citealt{SO2019}), or whether doing so may lead to biased cosmological constraints. To do so, we will compute the cluster abundances to be detected by these experiments both if the optimisation bias is neglected and if it is accounted for, comparing them appropriately.

\subsubsection{Experiment model}

We model our cluster number counts analysis as follows. We assume that an Internal Linear Combination (ILC; see, e.g., \citealt{Delabrouille2007}) is applied on its temperature frequency maps in order to produce a Compton-$y$ map covering a fraction $f_{\mathrm{sky}}$ of the sky. For \textit{Planck} we take $f_{\mathrm{sky}} = 0.62$ (e.g., \citealt{Ade2016}), whereas for SO we take $f_{\mathrm{sky}} = 0.4$ \citep{SO2019}. We then assume that galaxy clusters are detected by applying a matched filter on this Compton-$y$ map, which is previously subdivided appropriately into a number of tiles, maximising the signal-to-noise over angular size and sky location (i.e., three parameters) and imposing a selection threshold $q_{\mathrm{th}}$. For \textit{Planck} we take a selection threshold of $q_{\mathrm{th}} = 6$, which is equal to that used to construct the MMF3 cosmology sample used in the \textit{Planck} baseline analysis \citep{Ade2016}, whereas for SO we consider two selection thresholds, $q_{\mathrm{th}} = 5$ and $q_{\mathrm{th}} = 6$. We note that, as we show in Appendix \ref{appendix}, this ILC + matched filter approach is equivalent to a multi-frequency matched filter approach, which was followed, e.g., in the real \textit{Planck} analysis (see, e.g., \citealt{Planck2014XX}).

Construction of the matched filter requires as input a model for the signal to be detected, that is, a model for the cluster Compton-$y$ signal in our case, as well as the noise covariance. As the cluster model we assume the \citet{Arnaud2010} universal pressure profile, which integrated along the line-of-sight (see Eq. \ref{compton}) gives the cluster Compton-$y$ signal at given mass $(1-b) M_{500}$ and redshift $z$, where $1-b$ is the so-called `hydrostatic' mass bias. We use the best-fit parameters of \citet{Arnaud2010} and a mass bias of $1-b = 0.62$ (taken from \citealt{Planck2018I}). We write our model as  $y_{\mathrm{c}} (M_{500},z) = y_0 y_{\mathrm{t}} (\theta_{500} (M_{500},z))$, where $y_0$ is an amplitude parameter defined as the value of $y_{\mathrm{c}}$ at $R_{500}$, and $y_{\mathrm{t}}$ is an angular function that depends on mass and redshift only through the cluster's angular size, $\theta_{500} = R_{500}/ d_{\mathrm{A}}$, where $d_{\mathrm{A}}$ is the angular diameter distance to the cluster. Assuming this model for the cluster signal and the noise covariance, the matched filter noise at a given angular size $\theta_{500}$ is then given by

\begin{equation}\label{noisecluster}
    \sigma (\theta_{500}) =  \left[ \int d^2\bmath{l} \frac{|y_{\mathrm{t}} (\theta_{500};\bmath{l}) |^2}{N (l)} \right]^{-1/2},
\end{equation}
where $N(l)$ is the noise power spectrum. For \textit{Planck}, we take $N(l)$ to be the power spectrum of the Full Mission NILC Compton-$y$ map, which we estimate directly from the map, which is publicly available in the \textit{Planck} Legacy Archive\footnote{\texttt{pla.esac.esa.int}}. This estimated power spectrum includes the tSZ signal of galaxy clusters: this is analogous to what was done in the actual \textit{Planck} analysis, in which the clusters' tSZ signal was not subtracted from the frequency maps when estimating their power spectra. For SO, we take it to be the sum of the forecast baseline tSZ reconstruction noise power spectrum of \citet{SO2019} (namely, the solid orange curve of Fig. 36 of this paper) and the Compton-$y$ power spectrum due to halos up to masses of $M_{500} = 10^{14} M_{\odot}$, which we compute with the pubicly-available code \texttt{pysz}\footnote{\texttt{github.com/ryumakiya/pysz}}. This assumes that that the tSZ signal due to more massive halos will be identified and removed for noise estimation purposes. The mean signal-to-noise at given mass $M_{500}$ and redshift $z$ is then given by

 \begin{equation}\label{meansnr}
     \bar{q} (M_{500},z) = \frac{y_0 (M_{500},z)}{\sigma (\theta_{500} (M_{500},z))},
\end{equation}
where the dependency on the mass bias is left implicit. Note that Eq. (\ref{noisecluster}) is simply a Fourier space, cluster-specific version of Eq. (\ref{noise}). In order to evaluate the matched filter integral, we assume flat sky tiles of $1024 \times 1024$ pixels, with a pixel size of $0.8$\,arcmin for \textit{Planck} and of $0.2$\,arcmin for SO.

In order to link the mean signal-to-noise, given by Eq. (\ref{meansnr}), to $q_{\mathrm{opt}}$, we first assume a layer of log-normal intrinsic scatter. This introduces the variable $\ln q_{\mathrm{t}}$, which is given by a Gaussian centred on $\ln \bar{q} (M_{500},z)$ and with a standard deviation of $\sigma_\mathrm{\ln q} = 0.173$ (taken from \citealt{Ade2016}). We then consider a layer of `observational scatter' due to what is understood as noise in the matched filtering process, finally linking $q_{\mathrm{t}}$ to $q_{\mathrm{opt}}$. Motivated by our numerical results of Section \ref{mc_test}, we take this scatter to be Gaussian-distributed with unit variance and with a mean equal to

\begin{equation}\label{correction2}
\bar{q}_{\mathrm{opt}} = \sqrt{ q_{\mathrm{t}}^2 + f_{\mathrm{eff}}},
\end{equation}
where $f_{\mathrm{eff}}$ is the effective number of fitting parameters, and where we note that we have neglected the contribution due to the change in the variance (compare with Eq. \ref{correction2}), as it is small for $q \gtrsim 5$ (see Figure \ref{fig:prediction}). As discussed in Section \ref{mc_test}, for $q \gtrsim 6$, $f_{\mathrm{eff}} \simeq f = 3$. Thus, we take $f_{\mathrm{eff}} = 3$ when accounting for the bias; we take $f_{\mathrm{eff}} = 0$ when ignoring it. The differential mean number of clusters to be detected as a function of $q_{\mathrm{opt}}$ is then given by

\begin{equation}\label{barn}
   \frac{d \bar{N}}{d q_{\mathrm{opt}}}= 4 \pi f_{\mathrm{sky}} \int  P( q_{\mathrm{opt}} | M_{500},z)  \frac{d^4N}{d^3V dM_{500}} \frac{d^3V}{dz d^2\Omega} dM_{500}  dz\,.
\end{equation}
Here, $d^4N/(d^3V dM_{500})$ is the halo mass function, which we take to be that of \citet{Tinker2008}, $d^3V/(dz d^2\Omega)$ is the differential volume element, and $P( q_{\mathrm{opt}} | M_{500},z)$ is the probability for a cluster of mass $M_{500}$ and redshift $z$ to have an optimal signal-to-noise $q_{\mathrm{opt}}$, which is given by

\begin{equation}
    P(q_{\mathrm{opt}} | M_{500},z) = \int_0^{\infty}  P(q_{\mathrm{opt}}  | q_{\mathrm{t}} ) P(q_{\mathrm{t}} | M_{500},z) dq_{\mathrm{t}},
\end{equation}
where the first factor of the integrand is the Gaussian describing observational scatter, and the second one is the log-normal distribution accounting for intrinsic scatter. The mean number of clusters with $q_{\mathrm{opt}}$ between $q_{\mathrm{min}}$ and $q_{\mathrm{max}}$ is then given by

\begin{equation}\label{intnc}
    \bar{N} = \int_{q_{\mathrm{min}}}^{q_{\mathrm{max}}}  \frac{d \bar{N}}{d q_{\mathrm{opt}}} dq_{\mathrm{opt}}.
\end{equation}

We perform our calculations fixing all the cosmological parameters to their \textit{Planck} 2018 TT,TE,EE+lowE+lensing best-fit values \citep{Planck2018VI}.


\subsubsection{Results and discussion}

\begin{figure}
\centering
\includegraphics[width=0.5\textwidth]{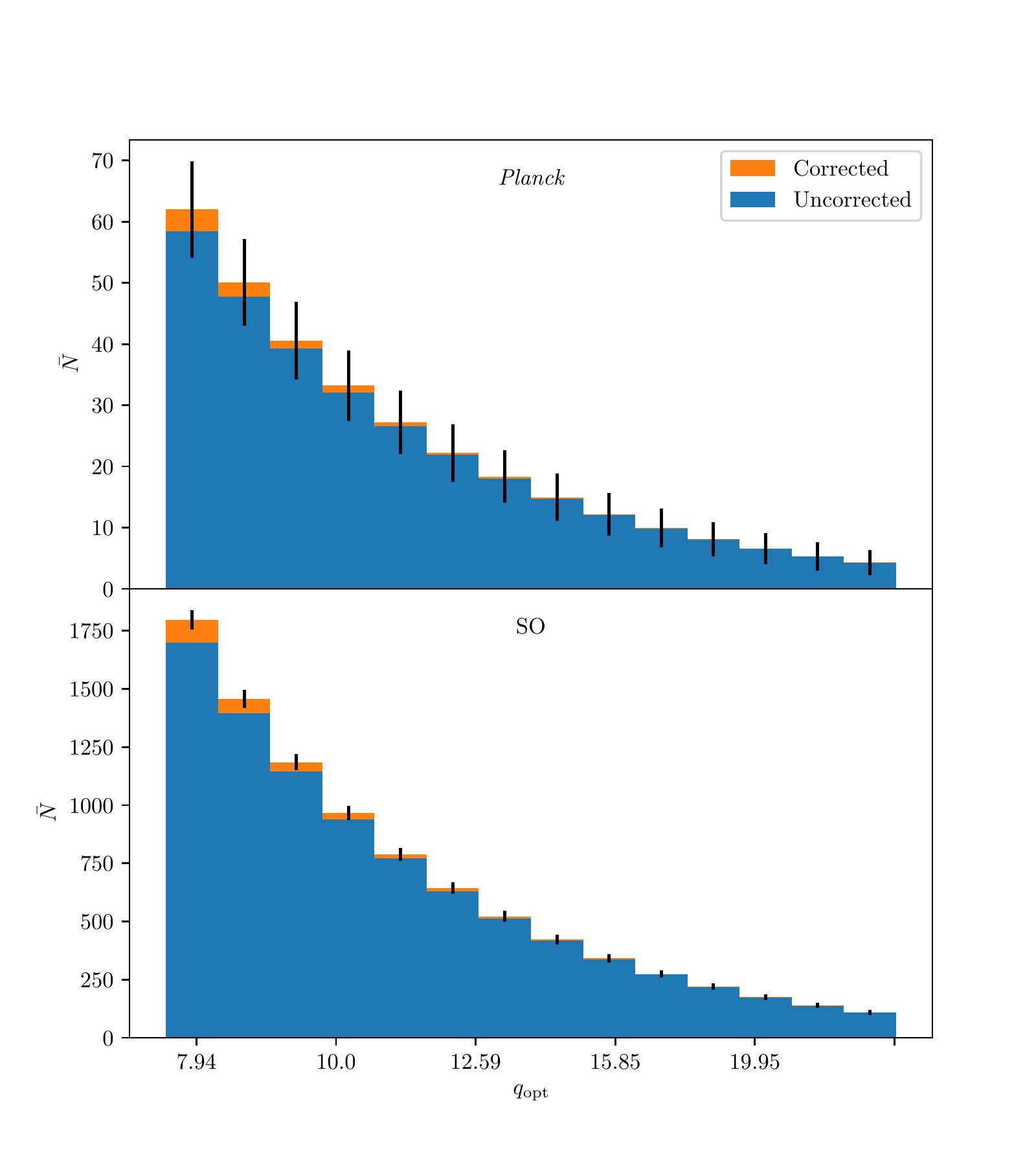}
\caption{Cluster abundances as a function of optimal signal-to-noise, $q_{\mathrm{opt}}$, for \textit{Planck} (upper panel) and SO (lower panel), both if the optimisation bias is neglected (blue bars) and if it is accounted for (orange bars, shown along the associated Poisson errors).}
\label{fig:histogram}
\end{figure}

Figure \ref{fig:histogram} shows the mean number of clusters as a function of $q_{\mathrm{opt}}$ for our reference \textit{Planck} and SO-like experiments both if the optimisation bias is taken into account (`corrected' orange bars) and if it is ignored by setting $f_{\mathrm{eff}}=0$ (`uncorrected' blue bars). Specifically, for each experiment, the mean number of clusters is computed using Eq. (\ref{intnc}) for 14 logarithmically-spaced bins between $q_{\mathrm{opt}} = 6$ and $q_{\mathrm{opt}} = 25$. As expected, in any bin more clusters are found if the optimisation bias is accounted for, since the cluster abundance decreases with $q_{\mathrm{opt}}$ and the optimisation bias boosts the signal-to-noise at given mass and redshift. Also as expected, the effect of the optimisation bias on the number counts is most significant at low signal-to-noise, quickly decreasing with increasing signal-to-noise.

Figure \ref{fig:histogram} also shows, as black error bars, the Poisson errors associated with the corrected number counts, which for each bin are given by $\Delta \bar{N}_{\mathrm{Poisson}} = \bar{N}^{1/2}$. They provide a measure of the change in the number counts that the analysis is sensitive to: if the change due to neglecting the optimisation bias is much smaller than the Poisson error, then it can be safely neglected, whereas if it is larger, it may have to be taken into account in order to avoid biasing the inference. It can be seen how for \textit{Planck} this change is smaller than the Poisson error across all the bins, whereas for SO it is larger for the lowest signal-to-noise bins. This implies that the impact of neglecting the optimisation bias in \textit{Planck} is probably negligible, but that this may not the case for SO.

Table \ref{numberclusters} shows the total mean number of clusters detected by each experiment, which is given by Eq. (\ref{intnc}) by setting its lower and upper integration limits to $q_{\mathrm{th}}$ and infinity, respectively. For \textit{Planck} we have imposed a selection threshold of $q_{\mathrm{th}} = 6$ (as in the real baseline analysis, see \citealt{Ade2016}), whereas for SO we have considered two selection thresholds,  $q_{\mathrm{th}} = 6$ and  $q_{\mathrm{th}} = 5$. As with Figure \ref{fig:histogram}, the number counts are given both if the optimisation bias is accounted for and if it is neglected; their difference and the associated Poisson errors are also shown. As expected from Figure \ref{fig:histogram}, while the change in the number counts for \textit{Planck} is well within the Poisson error, this is not the case for SO, especially if the selection threshold is lowered to $q_{\mathrm{th}} = 5$. This again implies that the optimisation bias probably has to be taken into account in SO, especially if a selection threshold as low as $q_{\mathrm{th}} = 5$ is chosen.


The number counts in Figure \ref{fig:histogram} and in Table \ref{numberclusters} were computed at fixed cosmological and scaling relation parameters. The uncertainty in the scaling relation parameters inherent to any real analysis means that the changes in the number counts shown in both Figure \ref{fig:histogram} and Table \ref{numberclusters} can thus be interpreted as an upper limit on the change that is discernible by the analysis and that can be (wrongly) attributed to cosmology. Indeed, if the tSZ-mass scaling relation is given enough freedom, the optimisation bias may be calibrated away to some extent\footnote{While the optimisation bias is not completely degenerate with a power law tSZ--mass scaling relation (the form that is usually assumed), as it does not have a power law behaviour, some degeneracy is expected.}. Quantifying the extent to which this can happen is beyond the scope of this paper. It could be done by generating a set of mock cluster samples (e.g., by sampling from the halo mass function and adding the appropriate scatter) and analysing them with a number counts likelihood (e.g., with the one used in \citealt{Zubeldia2019}) with and without the optimisation bias taken into account. One could then look at the differences in the derived constraints on the relevant cosmological parameters (e.g., $\Omega_{\mathrm{m}}$ and $\sigma_8$). We have decided not to pursue this avenue, as our conclusions would be dependent on how well the tSZ-mass scaling relation is calibrated, which is yet to be determined for SO, the experiment for which this full analysis would be interesting to perform. Furthermore, since the optimisation bias can be accounted for, and, indeed, implemented numerically in a very simple way (to lowest order, at least), we do not find it necessary to analyse its impact on our idealised experiments in any more detail. We simply recommend it to be considered in any future number counts analysis that makes use of $q_{\mathrm{opt}}$, or any equivalent quantity, as an observable.

\begin{table}
\centering
\caption{Mean number of clusters to be detected by our \textit{Planck} and SO-like experiments, with and without optimisation bias taken into account (first and second columns, respectively). Their difference is also shown (third column), along with the corresponding Poisson error (fourth column).}
\label{numberclusters}
\begin{tabular}{@{}lllll@{}}
\toprule
Experiment  & $\bar{N}$ (corr.)    & $\bar{N}$ (uncorr.)  & $\Delta \bar{N}$ & $\Delta \bar{N}_{\mathrm{Poisson}}$  \\
\midrule
\textit{Planck}  & 334.0 & 320.7  & 13.3 & 18.3 \\
SO ($q_{\mathrm{th}} = 6$) & 9400.9 & 9116.9 & 284.0 & 97.0 \\
SO ($q_{\mathrm{th}} = 5$) & 14837.6 & 14090.8 & 746.8 & 121.8 \\
\bottomrule
\end{tabular}
\end{table}

\section{Conclusion}\label{sec:conclusion}

In this paper, we have studied in detail the statistical properties of the matched filter optimal signal-to-noise, $q_{\mathrm{opt}}$, in the context of galaxy cluster tSZ detection and characterisation. In particular, we have shown that $q_{\mathrm{opt}}$ is, in general, a non-Gaussian variable with a standard deviation different from unity and an expected value that is biased high with respect to the true mean signal-to-noise, a bias that we have referred to as optimisation bias. We have shown that this bias arises from the fact that $q_{\mathrm{opt}}$ is obtained from an optimisation procedure on noisy data. As such, it is present even if the signal is modelled perfectly, no foregrounds are present, and the noise in the data is Gaussian. After presenting these arguments in a general way in Section \ref{sec:theory}, we have then used a set of MC mock observations in order to study numerically these effects of optimisation in the tSZ galaxy cluster context. We have found that, if optimisation is carried out over three parameters (an angular size and two location parameters), $q_{\mathrm{opt}}$ can be taken to be approximately Gaussian-distributed and with a standard deviation approximately equal to unity for $q_{\mathrm{opt}} \gtrsim 6$. Similarly, for $q_{\mathrm{opt}} \gtrsim 6$, the optimisation bias can be accounted for with a simple expression, $\bar{q}_{\mathrm{opt}} =  (q_{\mathrm{t}}^2 + f)^{1/2}$, where  $\bar{q}_{\mathrm{opt}}$ is the expected value of  $q_{\mathrm{opt}}$, $\bar{q}_{\mathrm{t}}$ is the expected value of the true signal-to-noise, and $f$ is the number of fitting parameters (three in our case). We stress that this correction, which we have refer to as the `$f$ correction', is accurate if `true' optimisation is achieved: if the signal-to-noise is only partially optimised by evaluating it on a coarse parameter grid, an additional, grid-dependent negative bias on the retrieved signal-to-noise will be induced (see Section \ref{sec:gridbias}).

In addition, we note that if in a given context the $f$ correction may be suspected not be accurate enough, or the standard deviation of $q_{\mathrm{opt}}$ may be significantly different from unity, this paper offers a route through which these aspects can be quantified, namely through the generation and analysis of a set of mock $q_{\mathrm{opt}}$ observations. This may be the case if low signal-to-noise clusters are considered, or if the parameters over which the optimisation is carried out are different from the ones we have considered, either in nature and/or in number. Indeed, while in this paper we have considered optimisation of the signal-to-noise over three parameters, in future experiments additional parameters could be considered, e.g.,
including extensions in the modelling of the cluster atmosphere \citep[e.g.,][]{Moser2021}, or
the cluster temperature \citep{Remazeilles2019, Remazeilles2020} in experiments sensitive to the relativistic SZ effect such as PICO \citep{Hanany2019}.

Our results have implications for any cluster number counts analysis that may use $q_{\mathrm{opt}}$ as a mass observable. In these analyses, a signal-to-noise threshold of around 6 is typically used as the selection criterion for the construction of the cluster sample. Thus, our results suggest that, while the non-Gaussianity of $q_{\mathrm{opt}}$ and the deviation of its standard deviation from unity are small and probably negligible, the optimisation bias, which is a $\sim 5$\% effect at signal-to-noise of $6$, may have a significant impact. In the past, analyses using both \textit{Planck} and ACT clusters have ignored this bias, even if optimisation over a number of parameters was carried out (three parameters in \textit{Planck} and two in ACT; \citealt{Hasselfield2013,Planck2014XX,Ade2016,Zubeldia2019}). Only the SPT analyses, in which the signal-to-noise was optimised over three parameters, have properly accounted for it using a correction equal to ours \citep{Vanderlinde2010,Bleem2015,Bocquet2018}, offering a rather heuristic justification for its use \citep{Vanderlinde2010}. Here we have quantified the relevance of the bias in \textit{Planck} and SO by computing its effect on the expected number counts across signal-to-noise. We have found it to be negligible for \textit{Planck} and potentially significant for SO, the latter being especially true if a precise mass--observable tSZ scaling relation is used in the analysis. We also expect the bias to be relevant for CMB-S4, which is forecast to find about an order of magnitude more clusters than SO \citep{Abazajian2016}.

This paper highlights the well-known fact that, though galaxy clusters are and will continue to be a very powerful cosmological probe from a statistical point of view, accurately determining the mass-observable relation is crucial for this statistical power to be realised in practice. For the specific case considered here, that of SZ surveys using matched filters, we have evidenced that, for upcoming surveys, accurately calibrating the mass--$q_{\mathrm{opt}}$ relation will probably require proper modelling of the optimisation bias. Here we have shown that this can be done in a numerically straightforward way. We thus encourage that the bias is duly accounted for in any future analysis.

Finally, although in this work we have focused on tSZ detection and characterisation of galaxy clusters, matched filters can also be used for cluster detection in photometric surveys (e.g., \citealt{Bellagamba2018}) and using jointly mm and X-ray observations \citep{Tarrio2018}, as well as for the confirmation of X-ray detected clusters using optical photometric data \citep{Klein2018}. They also have a variety of other applications, such as the detection of the CMB lensing signal of galaxy clusters \citep{Melin2015,Horowitz2019,Zubeldia2019} and voids \citep{Raghunathan2020}, of the moving lens effect \citep{Hotinli2021}, of line intensity mapping sources \citep{Schaan2021}, of large-scale structures such as voids in 21\,cm data \citep{White2017}, and of the ISW signal of large-scale structures \citep{Nadathur2016}. They are also widely used for gravitational wave detection (see, e.g., \citealt{Pitkin2011} for a review). All the theoretical insights of Section \ref{sec:theory}, including the non-Gaussianity of the signal-to-noise and the existence of a positive optimisation bias, are completely general and hold in any other applications of matched filters. We expect the general numerical trends presented in Section \ref{sec:cluster} (e.g., the non-Gaussianity and the optimisation bias both decreasing with increasing signal-to-noise) to apply in other contexts as well. Considering other uses of matched filters in any detail is beyond the scope of this paper, but we hope that the insights and results presented here may find useful applications in some of these other areas too.




\section*{Acknowledgements}

This work was supported by the ERC Consolidator Grant {\it CMBSPEC} (No.~725456). JC was furthermore supported by the Royal Society as a Royal Society University Research Fellow at the University of Manchester, UK.

\section*{Data Availability}

The data underlying this article will be shared on reasonable request to the corresponding author.



\bibliographystyle{mnras}
\bibliography{references} 




\appendix

\section{Equivalence between MMF and ILC+MF}\label{appendix}

When multi-frequency mm data is available, galaxy clusters are typically detected using a multi-frequency matched filter (MMF), which operates directly on the frequency maps (\citealt{Melin2006}; for an application on real data, see, e.g., \citealt{Planck2016xxvii}). Throughout this paper, however, we have considered a `single-frequency' matched filter acting on a Compton-$y$ map. In this appendix, we show that these two approaches are mathematically equivalent if the Compton-$y$ map is produced with an Internal Linear Combination (ILC; see, e.g., \citealt{Delabrouille2007}). Thus, the results presented in this work are also applicable to multi-frequency matched filters. We restrict our proof to the case in which the noise covariance is isotropic.

Consider a set of beam-deconvolved intensity maps at different frequencies, $\bmath{d}(\bmath{x})$, where the vector dimension of $\bmath{d}$ is equal to the number of frequency channels and where $\mathbfit{x}$ denotes angular position. We assume that  $\mathbfit{d}(\mathbfit{x})$ can be written as

\begin{equation}
    \mathbfit{d} (\mathbfit{x}) = \mathbfit{j}_{\nu} y(\mathbfit{x}) + \mathbfit{n} (\mathbfit{x}),
\end{equation}
where $\mathbfit{j}_{\nu}$ is the frequency dependency of the signal of interest (the tSZ effect in our case), $y(\mathbfit{x})$ is its spatial variation (the Compton-$y$ map in our case), and $\mathbfit{n} (\mathbfit{x})$ is some additive `noise', which also includes all other signals different from the component of interest (in our case, the CMB, the kSZ signal, etc.). Assuming the covariance of the noise to be isotropic, a Fourier-space ILC estimate of $y(\bmath{x})$ can be written as

\begin{equation}\label{ilc}
    \hat{y} (\mathbfit{l}) = \frac{\mathbfit{j}_{\nu}^T \mathbfss{N}^{-1} (l)  \mathbfit{d} (\mathbfit{l}) }{\mathbfit{j}_{\nu}^T \mathbfss{N}^{-1} (l) \mathbfit{j}_{\nu}},
\end{equation}
where $\mathbfss{N} (l)$ is the noise covariance matrix, and which has an associated variance equal to

\begin{equation}
    N (l) = \left[\mathbfit{j}_{\nu}^T \mathbfss{N}^{-1} (l) \mathbfit{j}_{\nu}\right]^{-1}.
\end{equation}
If we then assume that the signal $y(\mathbfit{l})$ can be written as $y(\mathbfit{l}) = y_0 y_{\mathrm{t}} (\mathbfit{l})$, where $y_0$ is an amplitude parameter and $y_{\mathrm{t}} (\mathbfit{l})$ is a spatial template, a matched filter estimate for $y_0$ can then be written as
\begin{equation}
    \hat{y}_0 =   \left[ \int d^2 \mathbfit{l} \frac{ |y_{\mathrm{t}} (\mathbfit{l})|^2 }{N(l)} \right]^{-1} \int d^2 \mathbfit{l} \frac{  y_{\mathrm{t}}^{\ast} (\mathbfit{l}) \hat{y} (\mathbfit{l}) }{N(l)}.
\end{equation}
This is the `single-frequency' matched filter that we have considered in this work. If we now substitute the ILC expression for $\hat{y} (\mathbfit{l})$ of Eq. (\ref{ilc}), we find

\begin{equation}
        \hat{y}_0 =   \left[ \int d^2 \mathbfit{l}  y_{\mathrm{t}}^{\ast} (\mathbfit{l})  \mathbfit{j}_{\nu}^T \mathbfss{N}^{-1} (l) \mathbfit{j}_{\nu} y_{\mathrm{t}} (\mathbfit{l})  \right] ^{-1}\int d^2 \mathbfit{l}  y_{\mathrm{t}}^{\ast} (\mathbfit{l})  \mathbfit{j}_{\nu}^T \mathbfss{N}^{-1} (l) \mathbfit{d} (\mathbfit{l}) .
\end{equation}
This is a multi-frequency matched filter, acting directly on the frequency maps $\mathbfit{d} (\mathbfit{l})$ (for comparison, see, e.g., \citealt{Melin2006}). Thus, it is equivalent to a `single-frequency' matched filter operating on an ILC estimate of the Compton-$y$ map.


\bsp	
\label{lastpage}
\end{document}